\documentstyle{mn}
\input psfig.sty
\begin{document}
\title[binary evolution \& blue stragglers]
{Effects of chemical composition and thermohaline mixing 
on the accreting components for low--mass close binaries: 
application to blue stragglers}
\author[Chen X. and Han, Z.]{Xuefei Chen$^{1,2}$\thanks{
xuefeichen717@hotmail.com} and Zhanwen Han$^{1}$ \\
$^1$National Astronomical Observatories/Yunnan Observatory, CAS, Kunming,
650011, P.R.China\\
$^2$Graduate School of the Chinese Academy of Sciences}
\maketitle

\begin{abstract}
Blue stragglers (BSs) are important objects in cluster populations 
because of their peculiar properties.
The colours and magnitudes of these objects are critical parameters
in population synthesis of the host cluster and 
may depend remarkably on BSs' surface composition.  
Observations show that some BSs are short-orbital-period binaries, 
which may be accounted for by mass transfer in low-mass binaries.
We therefore studied 
the effects of surface composition and thermohaline mixing 
caused by secular instability 
on the accreting components for low--mass binaries and 
applied the results on a short-orbital-period BS F190 in the old 
cluster M67.

We examine thermohaline mixing in a low-mass accreting-main-sequence star 
and find that, except the redistribution of composition under the surface,  
the mixing affects the accretor very little during Roch lobe overflow
unless thermohline mixing is treated as an instantaneous process.
A series of calculations are then carried out for low-mass binaries 
under different assumptions.
The results indicate no distinction in surface composition  
between the models with and without thermohaline mixing 
during Roche lobe overflow, 
but we still see the divergences of evolutionary tracks on 
Hertzsprung-Russell diagram and colour-magnitude diagram. 
The change of surface composition makes the gainer  
bluer and smaller than the ones with original surface composition 
while thermohaline mixing lessens the effect slightly.
If thermohaline mixing were to act instantaneously, 
the effect would be lessened more.
Our calculation shows that 
case A and case B mass transfer may produce BSs in short-
or relatively short-orbital-period binaries (including Algol systems),
and that  
CNO abundance abnormalities could be observed in these products.   
This is consistent with 
the results of Monte-Carlo simulations by previous studies. 

Our simulation of F190 shows that the primary's mass $M_{\rm 1i}$ 
of the appropriate models is located in 
the range of 1.40 to 1.45$M_{\odot}$ 
with initial mass ratio $q_{\rm i}=1.5$ 
and initial orbital period $P_{\rm i}=0.8$ days, 
indicating that case A is a more likely evolutionary channel 
than case B to form this object.  
The simulation also shows that it is very likely that 
F190 is still in a slow stage of mass transfer. 
As a consequence, obvious CNO abundance abnormalities 
should be observed for the object.
\end{abstract}


\section{Introduction}
Blue stragglers (BSs), which have stayed on the main sequence for a time
exceeding that expected from standard stellar evolution theory 
for their masses,
are important in population synthesis because of their peculiar properties. 
These objects lie above and blueward of the turn-off 
in the colour--magnitude diagram (CMD) of a cluster, 
may contribute remarkably spectral energy 
in the blue and ultraviolet, 
and affect the integrated spectrum of the host clusters
as they are bright and blue \cite{deng99}. 
The characters of BSs, i.e. luminosity, temperature, gravity etc., 
are relevant to their formation mechanism.
There are various possible origins for BSs in theory, i.e.  
close-binary evolution 
(mass transfer from a companion or coalescence of both companions), 
stellar collisions (single-single, binary-single and binary-binary), 
interior mixing, recent star formation etc. \cite{str93}.
Given the diversity of BSs within one cluster,
it is likely that more than one formation mechanisms play a role \cite{leo96}.
Observational evidence shows that 
binaries are at least important in some cluster BSs and in some field BSs
\cite{car01,leon96,ml92,pet84,str93}.

Since the binary mass-exchange hypothesis was originally advanced by McCrea 
\shortcite{mcc64}
to explain the BS phenomenon, 
a number of attempts have been made to test the hypothesis 
in open clusters \cite{col84,pol94,hur01}.
However there is a lack of detailed binary evolution calculations.
Monte-Carlo simulations \cite{col84,pol94} show that 
binary coalescence via case A evolution
(mass transfer begins when the primary is on the main sequence)
may be an important source of BSs in some clusters 
while case B evolution 
(mass transfer begins when the primary is in Hertzsprung gap) 
can only account for BSs in short-orbital-period binaries.
Meanwhile, case A may also produce BSs by stable mass transfer
as it does not always or immediately lead to a merger. 
The difficulty in verifying the binary mass transfer hypothesis is 
the lack of evidence for variations of radial velocities for most BSs. 
However some BSs have already been confirmed 
to be in binaries. A typical example is F190, 
a single-lined spectroscopic binary with a 4.2 days period \cite{ml92} 
in the old open cluster M67.
As well, IUE (International Ultraviolet Explorer) spectra \cite{land98}
provide evidence that F90 and F131 (in M67) 
are Algol-type mass transfer systems. 
With the improvement of observational means, 
more and more BSs are detected 
to have variations in radial velocity. 
It is therefore necessary to study the hypothesis in detail.  
As more hydrogen is mixed into the center of an accretor 
which is still on the main sequence at the onset of Roche lobe overflow(RLOF),
the accreting component goes upwards along the 
main sequence in response to accretion 
and its time on the main sequence is extended.
When the mass of the accretor is more massive than 
the corresponding cluster turn-off mass
at the age of the cluster, 
it may be recognized as a BS
and show element contamination from the primary on the surface. 
The element contamination will effect observational characteristics of the star
and it is dependent on the details of the accretion process.

Evolution of close binaries has been well studied 
over the last decades (see a review of van den Heuvel \shortcite{heu94}
and some recent papers by de loore \& Vanbeveren \shortcite{del95},
Marks \& Sarna \shortcite{mar98}, Han et al.\shortcite{han00},
Hurley, Tout \& Pols \shortcite{hur02}
and Nelson \& Eggleton \shortcite{nel01} etc.).  
Many of these works are related to the secondary
(initially lower mass component in a binary), 
where accretion mode and accretion rate are concerned on. 
Since the accreting matter 
may be originally in the convective core of the primary, 
thermohaline mixing,
which results from accreting material with a higher molecular weight than 
the surface layers, 
was first introduced by Ulrich \shortcite{ulr72} to describe this effect
(see also Kippenhanhn, Ruschenplatt \& Thomas\shortcite{kip80}). 
It was treated as an instantaneous process in intermediate-mass and massive 
close binaries for its short time scale in these systems by some authors
\cite{hel84,del94,del95}. 
As well, the details of accretion process during RLOF 
were ignored in these studies, and 
we have no knowledge of the effect of composition of secondaries during RLOF.
Wellstein, Langer \& Braun \shortcite{wlb01} treated thermohaline mixing 
in a time-dependent way in massive binary evolution and argued that 
this is important 
as both thermohaline mixing and accretion occur on a thermal timescale. 
In this paper, we concentrate on the behavior of secondaries in low-mass 
binaries during or just ceasing RLOF, applying the results on 
short-orbital-period BSs.
The computations are introduced in section 2 and  
thermohaline mixing for low-mass close binaries is examined in section 3.
The results are shown in section 4.
In section 5, we show a best model of F190 by simulation 
in a complete parameter space.   
Finally we give our conclusions, discussions and outlook on the work. 

\section{computations}   
We use the stellar evolution code 
devised by Eggleton \shortcite{egg71,egg72,egg73} and updated 
with the latest physics over the last three decades \cite{han94,pol95,pol98}.
The ROLF is included via the boundary condition

\begin{equation}
$${{\rm d}m \over {\rm d}t}$$=C \cdot {\rm Max}[0,($${r_{\rm star}\over r_{\rm lobe}}$$-1)^3],
\end{equation}
when we follow the evolution of the mass donor. 
Here ${\rm d}m/{\rm d}t$ is the mass loss rate of the primary,
$r_{\rm star}$ is the radius of the star, $r_{\rm lobe}$ is the radius 
of its Roche lobe and C is a constant. 
Here we take $C=500{M_\odot}{\rm yr^{-1}}$ 
so that RLOF can proceed steadily and the lobe-filling star overfills its 
Roche lobe as necessary but never overfills it by much, 
i.e a transfer rate of $5\times10^{-7}M_\odot {\rm yr^{-1}}$ 
corresponding to an overfill of 0.1 per cent. 
When following the evolution of the primary, 
we store its mass-loss history 
as an input in subsequent calculations of the secondary, 
including the age when RLOF begins, mass loss rate, 
and the composition of the lost matter.
The calculation of the secondary is stopped 
as it overfills its Roche lobe
and the system becomes a contact binary. 
Contact binaries are beyond our considerations here
for there are many uncertainties during contact phase.
Merger models will be studied in another paper.
 
The accreting matter is assumed to be deposited onto 
the surface of the secondary with zero falling velocity 
and distributed homogeneously all over the outer layers. 
The change of chemical composition on the secondary's surface 
caused by the accreting matter is 
\begin{equation}
($${\partial X_i \over \partial t }$$)=$${(\partial M /\partial t)\over (\partial M /\partial t){\rm d}t+M_{\rm s}}$$(X_{i{\rm a}}-X_{i{\rm s}})
\end {equation}
where $\partial M /\partial t$ is the mass accretion rate, 
$X_{i{\rm a}}$  and $X_{i{\rm s}}$ are element abundances of 
the accreting matter and of the secondary's surface for species $i$, 
respectively.
$M_{\rm s}$ is the mass of outer most layer of the secondary.
$M_{\rm s}$ will change with the moving of non-Lagrangian mesh as well as
the chosen resolution, 
but it is so small ($\sim 10^{-9}-10^{-12} M_{\odot}$) in comparison with 
$(\partial M /\partial t){\rm d}t$
($\sim 10^{-3}-10^{-5} M_{\odot}$) during RLOF
that we can ignore the effect of various $M_{\rm s}$ on element abundances.
Before and after RLOF,
we get $\partial X_{\rm i} /\partial t =0$ from the equation, 
which is reasonable in the absence of mixing.  

The opacity table is from OPAL \cite{ir96} with solar composition 
and from Alexander \& Ferguson \shortcite{af94a,af94b} 
in our calculations. 
Only stable RLOFs are considered in our calculations,
under which the accreting component may 
completely accrete the matter lost by the primary 
as mass transfer is very slow in this case
($\sim 10^{-9}M_\odot {\rm yr^{-1}}$, see section 4).
The mass and angular momentum of the systems are
therefore assumed to be conservative.
Convective overshooting has little effect 
in the mass range(1.0--1.6$M_{\odot}$) we considered 
\cite{ch03}. 
We, therefore, have not considered it in our calculations. 

\section{Secular Instability and Themohaline mixing}
During mass transfer in a close binary, 
the accreting matter may be originally in the convective core of the donor. 
It is rich in helium relative to the gainer's surface 
and has a higher mean molecular weight,
which will cause secular instability during or after accretion. 
Thermohaline mixing will occur in this case.  
Kippehahn, Ruschenplatt and Thomas \shortcite{kip80} 
estimated the time scale of the mixing for some massive binaries,
in which effects of radiation pressure and degeneracy were considered:
\begin{equation}
\tau_{\rm diff} \approx $${\overline{\mu} \over \Delta{\mu}}$$ $${{\delta}^2(\nabla_{\rm ad}-\nabla)\over 12{\phi}(1-\beta){\nabla_{\rm ad}}}$$$${W^3 \over H_{\rm p}lc},
\end{equation}
here $\nabla_{\rm ad}$, $\nabla$ and $H_{\rm p}$ have their usual meanings, 
$\overline{\mu}$ is the mean value of $\mu$ in the region of interest, 
$W$ is its distance in depth,   
$\beta$ is the ratio of gas pressure to total pressure, and
$l={1\over{\kappa}{\varrho}}$ and $c$ are the mean free path and 
velocity of a photon, respectively. $\delta $ and  $\phi$ are defined below :

\begin{equation}
\delta =-($${\partial {\rm ln}\varrho \over \partial {\rm ln} T }$$)_{P,\mu},
\phi =($${\partial {\rm ln}\varrho \over \partial {\rm ln}\mu}$$)_{P,T}.
\end{equation}

A massive-main-sequence star ($15M_\odot$) 
with a helium envelope ($1M_\odot$) was examined in that paper. 
Roughly estimated from a discontinuous model, 
in which exists a jump from the region of lower mean molecular weight 
to a higher one, the star has $\tau_{\rm diff} =1.7 \times 10^4 {\rm yrs}$
if diffusion extends over the whole radiative envelope.
Ulrich \shortcite{ulr72} obtained a time scale of 400 yrs in a similar case.
Thermohaline mixing was then treated as an instantaneous process 
for intermediate-mass and massive close binaries by some authors 
in previous studies \cite{del94,del95}. 
However the actual time is longer than $1.7 \times 10^4 {\rm yrs}$
for some reasons, e.g., ${\Delta \mu}$ decreases gradually 
while undergoing thermohaline mixing.

\begin{figure}
\centerline{\psfig{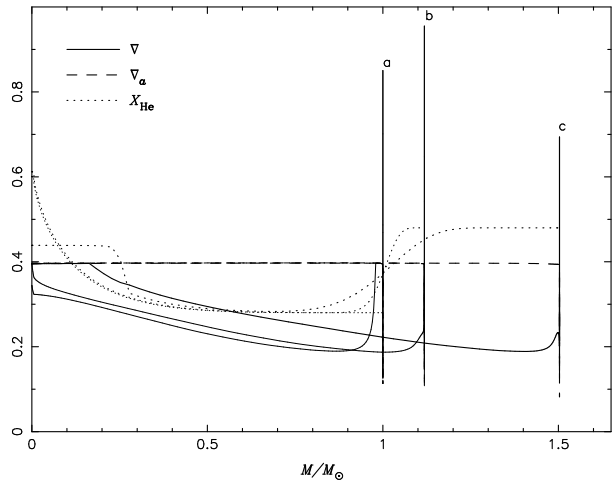}}
\caption{Profiles of adiabatic temperature gradient, 
real temperature gradient and helium composition at different masses
that the accretor reaches.
a, b and c represent the mass of 1.00, 1.12 and $1.50M_\odot$, respectively.
Thermohaline mixing has not been included here.}
\label{grad}
\end{figure}

A binary 1.26$M_\odot$ + 1.00$M_\odot$ 
with an initial orbital period of 1.0 day 
is considered here to examine the effect of thermohaline mixing on
low-mass binaries.
In a low--mass binary, 
He-rich matter is usually accreted onto the surface of the secondary
during the last phase of RLOF and the He-rich phenomenon is not very severe 
in the accreting matter.
The hydrogen mass fraction of the accreting matter is therefore assumed to be 
$X=0.5$ and helium mass fraction to be $Y=0.48$, 
which is an extreme case for low--mass binaries since  
hydrogen abundance usually has not reached such a low value.
Another remarkable feature for low--mass binaries is the existence of 
convective envelope, which may mix the accreting 
matter into the whole convective envelope very rapidly, 
minimizing thermohaline mixing in this region.
We therefore show the profile of thermodynamical quantities 
$\nabla_{\rm ad}$ and $\nabla$ during RLOF in figure \ref{grad}
to see the change of surface convective region caused 
by the increase of the mass.
a, b and c represent the mass of the secondary 
at about 1.00, 1.12 and $1.50M_\odot$, respectively.
The helium profile at corresponding masses is also shown in the figure.
Thermohaline mixing has not been included in the figure yet.

\begin{figure}
\centerline{\psfig{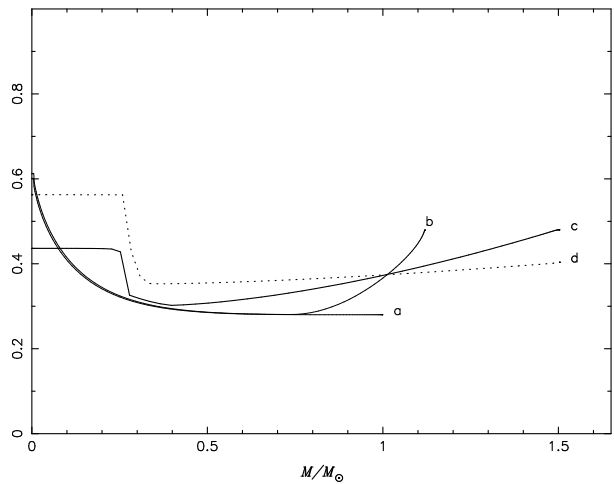}}
\caption{Helium profiles at different masses that the accretor reaches, 
themohaline mixing is added.
a, b and c represent the mass of 1.00, 1.12 and $1.50M_\odot$, respectively.
Dotted line represents the profile 
at an age of $\sim 5.3 \times 10^9 {\rm yrs}$, 
$ 5 \times 10^8 {\rm yrs}$ older than that of line c,
where the accretor just stops its accretion from the mass donor.}
\label{he1}
\end{figure}

\begin{table*}
\begin{minipage}{15cm}
\caption{Data for a ${1.5M_\odot}$ star with a He-rich envelope.
Radius, pressure scale height, photon mean free path, 
temperature gradient and the thermodynamic quantities
${1-\beta}$, ${\nabla_{ad}}$,${\delta}^2/\phi$ and $\overline{\mu}$ 
are given for different regions.}
\begin{tabular}{c|ccccccccc}
\hline
  location &$M_{r}/M_{\odot}$ &$r/10^{10}{\rm cm}$ &$H_{p}/10^8{\rm cm}$ 
  &$l/{\rm cm}$ &$\nabla$ &$1-\beta $
  &$\nabla_{\rm ad}$ &$\delta^2/\phi$ &$\overline{\mu}$ \\
\hline
 core     &0.164&1.020&90.4&0.0177&0.396&0.0015&0.396&1.0061&1.529\\
 boundary &0.692&2.000&64.0&0.0328&0.264&0.0011&0.397&1.0043&1.295\\
 surface  &1.503&8.303&1.8 &$2.7\times 10^7$&0.125&0.0053&0.087&1.0213&1.607\\

\hline 
\label{1}
\end{tabular}
\end{minipage}
\end{table*}
        
We find in figure \ref{grad} that 
the surface convective region is very small and
disappears quickly after a little matter is accreted, 
which cause He-rich matter pile up on the surface. 
In figure \ref{grad}, the helium profiles are smoothed due to 
numerical diffusion owing to the moving non-Lagrangin mesh.
A higher mesh resolution may improve it.

We stop the calculation when $0.5M_\odot$ matter is accreted. 
In Table \ref{1}, 
we show some thermodynamic quantities from the last model 
at the boundary of convective core, 
at which the mean molecular weight begins to increase outward 
(boundary of interest) and at the surface.
We obtained $\tau_{\rm diff} \approx 3.46 \times 10^8 {\rm yrs}$ 
from equation(3) and Table \ref{1}.
It is more than a quarter of the remaining 
main sequence time of the star after accretion 
($\sim 1.35 \times 10^9 {\rm yrs}$),
which means that we can observe He-rich phenomenon 
for about at least a quarter of the life of the star left on the main sequence 
after accretion.

We now examine the effect via detailed calculations 
where thermohaline mixing is included as a diffusion process. 
The diffusion coefficient is defined as 
(see Kippehahn, Ruschenplatt \& Thomas \shortcite{kip80})  
\begin{equation}
D=\alpha $${H_{\rm p}L^2 \over (\nabla_{\rm ad}-\nabla) \tau_{\rm KH} }$$ |$${\rm d} \mu \over {\rm d} r$$| $$1 \over \overline \mu$$=\alpha \cdot $$12(1-\beta){H_{\rm p}lc\nabla_{\rm ad}\phi \over (\nabla_{\rm ad}-\nabla) {\delta}^2}$$ |$${\rm d} \mu \over {\rm d} r$$| $$1 \over \overline \mu $$,
\end{equation}
where $\tau_{\rm KH}$ and $L$ can be considered as 
the Kelvin-Helmholtz time scale of perturbed blob and its radius, 
respectively. 
$\alpha$ is an efficiency parameter of order unity and can be increased  to 
strengthen this mixing. 
The same model as above is calculated ($\alpha=1.00$)
and helium profiles at corresponding masses are shown in
figure \ref{he1}. The dotted line in the figure represents helium profile 
at an age of $\sim 5.3 \times 10^9 {\rm yrs}$, 
where the star has evolved for some time after RLOF. 

\begin{figure}
\centerline{\psfig{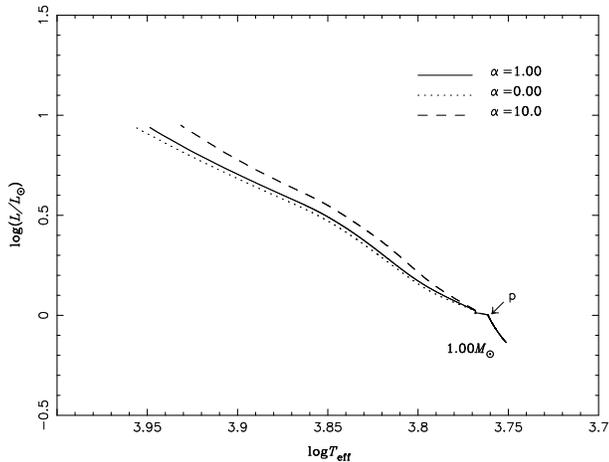}}
\caption{Evolutionary tracks of the secondaries during RLOF 
for models with and without thermohaline mixing.}
\label{hrd}
\end{figure}

We find  obvious differences in the helium profile 
between figure \ref{grad} and figure \ref{he1}, 
which cause the final discrepancy of evolutionary tracks (figure \ref{hrd}).
For comparison, the evolutionary track as $\alpha=10.0$,
which is very different from the case of $\alpha=1.00$, 
is also presented in figure \ref{hrd}.
The turn-off (point p) at the onset of RLOF in the figure 
results from a sudden change of surface composition. 
After RLOF, He-rich matter mixes inward quickly at first.
However the mixing becomes slower and slower 
because of the decrease of $\Delta {\mu}$ 
and increase of $W$.
The age at point c is about $4.8\times 10^9 {\rm yrs}$, 
$5\times 10^8 {\rm yrs}$ younger than that of point d, 
however,
we still see that $\mu$ slightly increases outwards in the envelope.
It means that the real time-scale of the mixing is much longer 
than the estimated value before.

For the reasons above, we prefer to a time-dependent way rather than 
an instantaneous process in treating thermohaline mixing 
in low-mass close binaries, especially when we are interested in  
the surface composition of the gainers.
As convective time scale is much shorter than that of thermohaline mixing,
we just consider the effect in radiative regions 
when we include thermohaline mixing.

\section{Results}
As examples, we calculate a series of low--mass binaries to examine 
the effects of surface composition and thermohaline mixing 
on these objects.
The initial mass of the primary  $M_{\rm 1i}$ has three values,
$1.0$, $1.26$ and $1.6M_\odot$, in roughly equal intervals in log$M_{\rm 1i}$. 
The initial mass ratio $q_{\rm i}=M_1/M_2$ of the systems is 
1.1, 1.5 and 2.0.
Case B evolution is simulated here 
and RLOF begins at early Hertzsprung-gap 
\cite{han00,ch02,ch03}.
The models are calculated under four different assumptions:
(a) the surface composition is assumed to remain unchanged during accretion;
(b) chemical composition of the accreting matter is included via equation (2);
(c) thermohaline mixing included based on (b), $\alpha =1.00$;
(d) similar to (c), but for $\alpha =10^5$, 
which can be considered as an instantaneous assumption. 
Accretion rate and element abundances of the accreting matter are determined 
by the primaries.

\begin{figure}
\centerline{\psfig{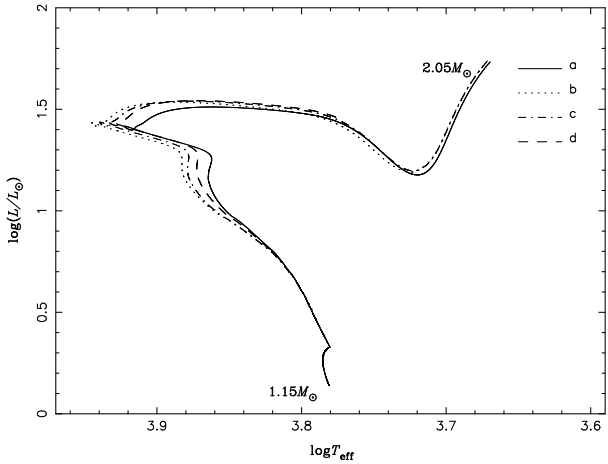}}
\caption{Evolutionary tracks of the secondary (evolving up from the bottom) 
 under different assumptions
for a binary of $1.26+1.15M_\odot$($q_{\rm i}=1.1$) 
with an initial orbital period of 0.943 days.}
\label{1261p1}
\end{figure}

\begin{figure}
\centerline{\psfig{figure=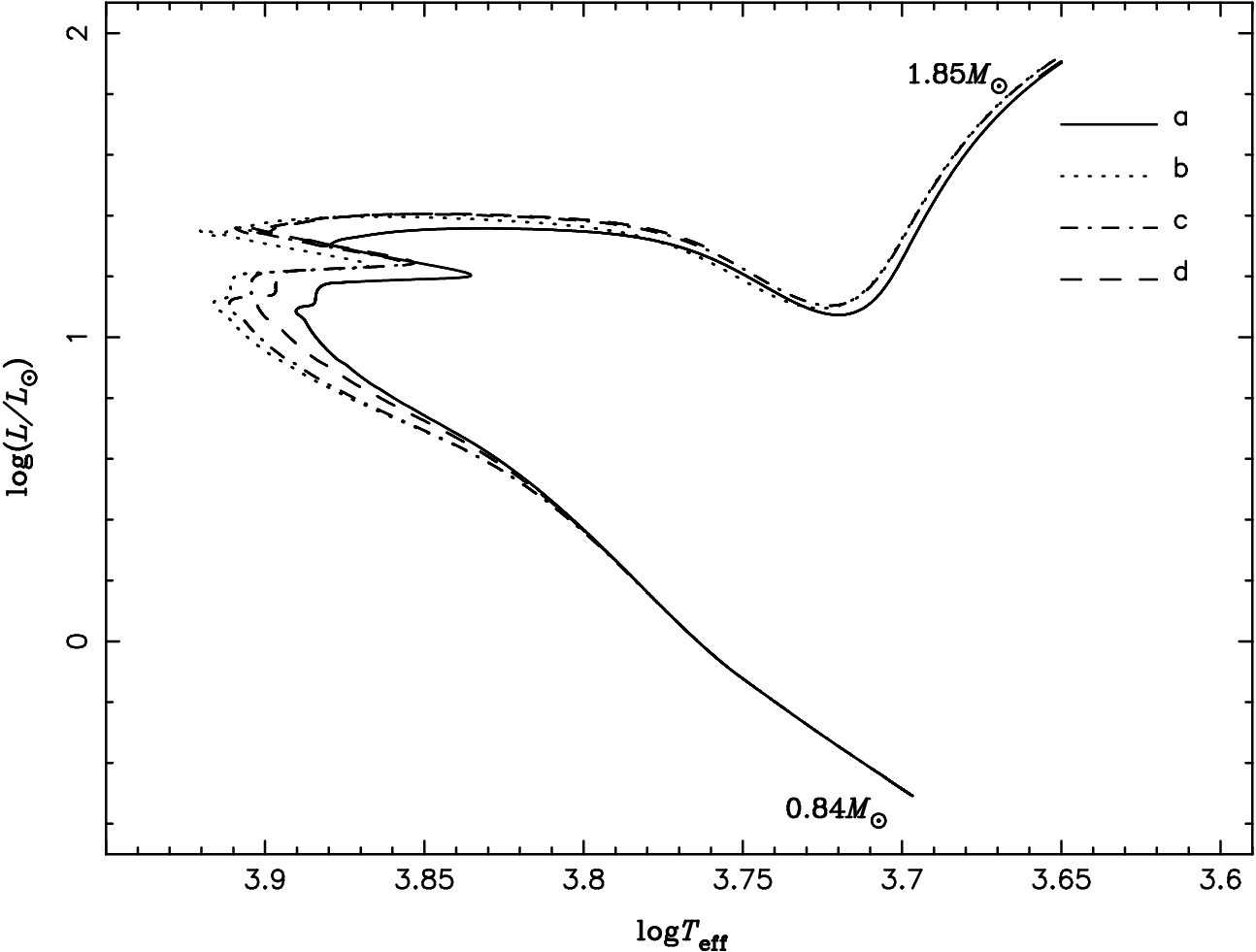,width=8cm,angle=270}}
\caption{Similar to figure \ref{1261p1}, but for a binary of 
$1.26+0.84M_\odot$($q_{\rm i}=1.5$) 
with an initial orbital period of 0.911 days.}
\label{1261p5}
\end{figure}

Evolutionary tracks for models with the primary mass $1.26M_\odot$ 
are shown in figures \ref{1261p1}, \ref{1261p5} and \ref {1262p0}. 
We find obvious divergences of evolutionary loci in the three figures.
The models under assumption (a) stay on the right 
and the ones with assumption (b) lie on the left.
The dash-dotted lines, much closer to that of models (b)  
represent the ones with thermohaline mixing included ($\alpha =1.0$)
and the dashed lines are for $\alpha =10^5$. 
From the three figures we see that, 
the change of surface composition, 
mainly the enhancement of helium abundance, makes the gainer hotter (or bluer),
thermohaline mixing just slightly lesses the effect while  
the instantaneous mixing ($\alpha=10^5$) may lessen the effect by much. 
All these finally cause corresponding divergences on CMD
(figure \ref{mbv126}). 

At the onset of RLOF, 
surface convective regions exist in the secondaries, however, 
the accreting matter comes from the surface of the primaries
and has a similar composition to that of the secondaries' surface,
the evolutionary loci then show no differences in the three figures. 
When the composition of the accreting matter becomes quite different 
at the last phase of RLOF,
surface convective regions of secondaries have already disappeared.
Then some He-rich matter is piled up on the surface of the accretors. 
We see in figure \ref{he1} that 
thermohaline mixing has no effect on surface element abundances during
RLOF, but it may redistribute the chemical composition in the interior
(under the outermost layers) 
and affect the evolutionary tracks (figures \ref{hrd} to \ref{1262p0}).

\begin{figure}
\centerline{\psfig{figure=1262p0.ps,width=8cm,angle=270}}
\caption{Similar to figure \ref{1261p1}, but for a binary 
of $1.26+0.84M_\odot$($q_{\rm i}=2.0$) 
with an initial orbital period of 0.878 days. }
\label{1262p0}
\end{figure}

\begin{figure}
\centerline{\psfig{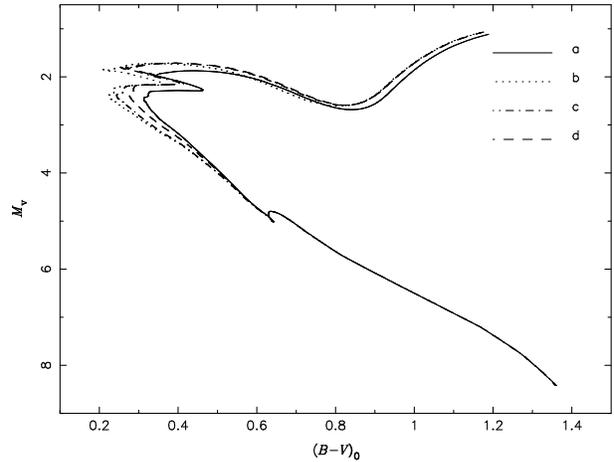}}
\caption{Evolutionary tracks of the secondary  
under different assumptions 
on color-magnitude diagram for a binary of $1.26+0.63M_\odot$ ($q_{\rm i}=2.0$)
with an initial orbital period of 0.878 days.}
\label{mbv126}
\end{figure}

\begin{figure}
\centerline{\psfig{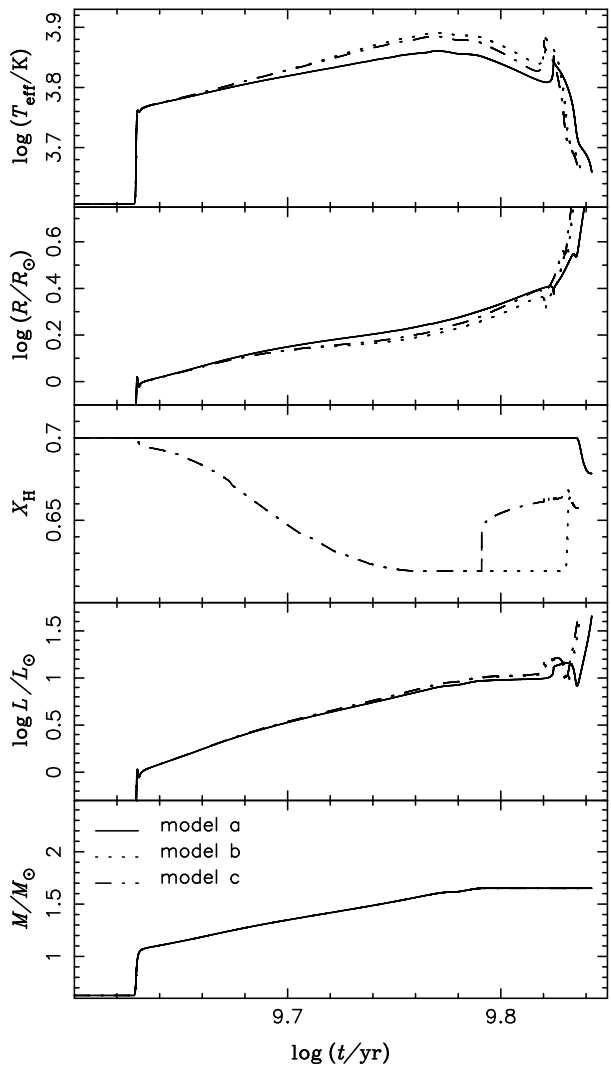}}
\caption{Some surface parameters change with  age for a binary of
$1.26+0.63M_\odot$ ($q_{\rm i}=2.0$) 
with an initial orbital period of 0.878 days.}
\label{1262p0a}
\end{figure}

We show some surface parameters of a binary with masses $1.26+0.63M_\odot$ 
and with an initial orbital period of $0.878$ days in figure \ref{1262p0a}.
The divergences of these parameters appear at 
the age ${\rm log}t{\rm (yrs)}=9.66$ 
when 
the surface hydrogen mass fraction has decreased obviously.
In the figure we see that,
with the decreasing of surface hydrogen, 
the star contracts slightly (which makes a increase of surface gravity)
compared to the one with original surface composition. 
As well, the surface effective temperature has increased.  
The change of both temperature and radius (or gravity) 
leads to variations of colour indices of the star,
which finally causes the change of position of the star on CMD 
(figure \ref{mbv126}).
Thermohaline mixing lessens the effect,
but the lessening is slight during RLOF 
unless it were to act instantaneously, i.e. $\alpha =10^5$.
We also see that 
little distinction in surface hydrogen mass fraction $X_{\rm H}$  
exists between the models with and without thermohaline mixing 
before log$t({\rm yrs})=9.79$ in figure \ref{1262p0a}.
However there is a jump in $X_{\rm H}$
near the end of the main sequence, 
corresponding  to the end of RLOF -- 
no matter is accreted during this time,  
and thermohaline mixing makes surface hydrogen increase again.

There are two interesting turn-offs on the evolutionary tracks in figure
\ref{1262p0}. 
One is before the divergence of the loci (${\rm p_1}$)
and the other is near the end of main sequence (${\rm p_2}$). 
The two points
correspond to two abrupt changes of mass accretion rate during RLOF. 
The first is a sudden increase of mass accretion rate 
at the onset of RLOF, 
the mass increases from 0.63 to 1.1$M_{\odot}$ in a very short time, 
and thermal adjustment of the star 
results in the first turn-off in the figure. 
The other turn-off is caused by the decrease of the mass accretion rate 
during the last phase of RLOF.

\begin{figure}
\centerline{\psfig{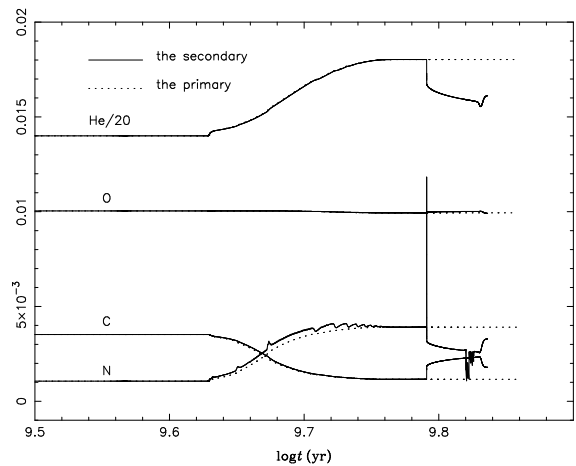}}
\caption{Helium and CNO surface abundances 
change with age for a binary of
$1.26+0.63M_\odot$ ($q_{\rm i}=2.0$) 
with an initial orbital period of 0.878 days.
Note the helium composition is divided by 20 in the figure.}
\label{CNO}
\end{figure}

\begin{figure}
\centerline{\psfig{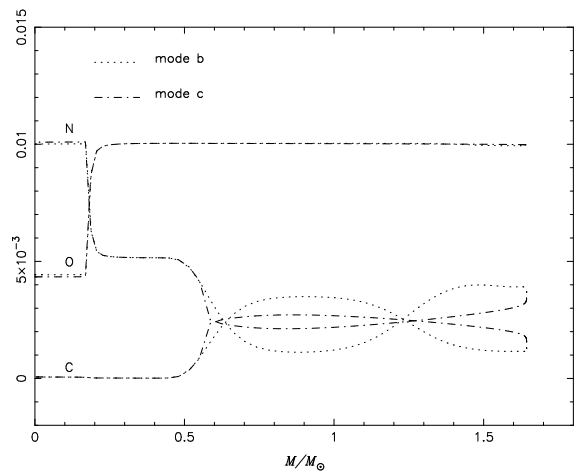}}
\caption{The profiles of CNO abundances of the accretor 
at the mass of about $1.64M_\odot$. The initial parameters of the system are
$1.26+0.63M_\odot$ ($q_{\rm i}=2.0$) 
with an initial orbital period of 0.878 days.}
\label{str-CNO}
\end{figure}

The variations of surface CNO abundances of the secondary for a binary 
$1.26+0.63M_\odot$ ($q_{\rm i}=2.0$) 
with an initial orbital period of 0.878 days
under assumption (c) are shown in figure \ref{CNO}. 
It is not distinctly different from the one under assumption (b) 
except C increases and N decreases at ${\rm log}t({\rm yrs})=9.79$, 
which is caused by thermohaline mixing taking effect 
on the surface abundances when accretion stops during this time.
We find obvious changes of surface CNO abundances of the secondary 
during and after RLOF in the figure.  
The changes of N (increased) and C (decreased) begin at 
${\rm log}t({\rm yrs})=9.63$ when the secondary mass is near $1.1M_\odot$.
The surface convective region has already disappeared or will disappear soon
for a star with this mass (see figure \ref{grad}).
With more matter lost from the primary, 
the change becomes more distinct -- 
C is reduced from 0.0035 to 0.001 while N increases from 0.001 to 0.004 
at ${\rm log}t({\rm yrs})=9.72$. 
Even after RLOF,
CNO abundances also show obvious abnormalities -- 
N is enhanced by nearly 150 per cent and C is decreased by nearly 50 per cent.
It means that if a BS is formed by binary mass transfer, 
whether it is an Algol system or not, 
we should observe clear indication of some surface contamination by 
processed material spilled over from an evolved companion.
CNO abundances of the accreting matter (from the primary) 
are also shown in the figure to check our code.
The N discrepancy comes from the precision of $X+Y$, 
for N abundance is calculated by 
$X_{\rm N}=1.0-X-Y-X_{\rm C}-X_{\rm O}-X_{\rm Ne}-X_{\rm Mg}-X_{\rm Si}-X_{\rm Fe}$
in our code.   
We cannot distinguish the models between assumption (b) and (c)
via surface CNO abundances during RLOF
as there is almost no difference in surface CNO composition between  
the models.

In order to examine the effect of thermohaline mixing, 
we show the profiles of CNO abundances in figure \ref{str-CNO}
when the secondary reaches $1.64M_\odot$. 
We see in the figure that 
the accreting matter has already mixed with the matter
under the outer most envelope, 
as leads the change of overall stellar properties
(figure \ref{1262p0a}). 
    
\begin{figure}
\centerline{\psfig{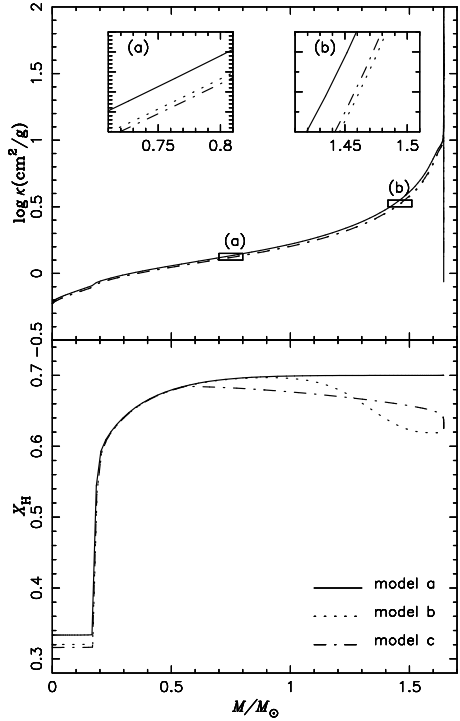}}
\caption{The profiles of opacity as the gainer reaches $1.64M_\odot$  
for the binary $1.26+0.63M_\odot$ with $P_i=0.878$ days.}
\label{fk2}
\end{figure}

The variations of chemical composition first result in  
the change of opacity, then other thermodynamic quantities,
i.e. temperature, entropy, pressure ect.. 
Figure \ref{fk2} shows the distribution of opacity 
for the accretor when it reaches 1.65$M_{\odot}$ for the system above. 
In the figure, log$\kappa$ of model (a) has already been distinguished from 
those of model (b) and (c),
but the effect of thermohaline mixing is not very remarkable -- 
the difference on log$\kappa$ between models (b) and (c) 
can only be seen by enlarging some parts of the figure.
We may clearly see the influence of hydrogen mass fraction on opacity 
in the figure -- larger hydrogen mass fraction results in larger opacity
(it might be invalid at elsewhere as opacity also depends on temperature). 
The change of opacity mainly affects on the effective temperature, 
which can be seen in figure \ref{1262p0a}--
model (b) has a obviously higher surface effective temperature 
than model (a), while model(c) reduces this difference slightly.  

As mentioned in section 1, 
a main-sequence accretor will go upwards along the main sequence 
in response to accretion and its main-sequence life will extend 
because hydrogen-rich matter mixes into the center. 
Here convective core is necessary 
to mix hydrogen-rich matter into the center.
For a low-mass accretor without convective core at the onset of 
RLOF, central hydrogen does not increase until a convective core develops 
with mass increasing.      
After a convective core develops, 
further accretion will lead hydrogen-rich matter 
be involved in the convective core, 
increase the central hydrogen mass fraction and finally extend its
life on the main sequence. 

\begin{table}
  \caption{System parameters when the secondaries 
           leave off the main sequence. * represents that 
           RLOF of the binary has not ceased yet at that time. 
           The bottom part is for our simulation of F190.
           All the binaries are calculated under assumption (c).}
   \label{end-MS}
   \begin{tabular}{ccccccccc}
\hline
 $\scriptstyle M_{\rm 1i}$ & $\scriptstyle q_{\rm i}$ &
   $\scriptstyle P_{\rm i}$ &
   $\scriptstyle M_{\rm 2}$ & $\scriptstyle P$ & 
   $\scriptstyle t$& \\
 $\scriptstyle (M_\odot)$ & &
   $\scriptstyle ({\rm days})$ &  
   $\scriptstyle (M_\odot)$ & 
   $\scriptstyle ({\rm days})$&
   $\scriptstyle ({\rm yrs})$ &\\
\hline
1.00& 1.1& 0.590& 1.3802& 1.099& 1.22E+10& *\\
1.00& 1.5& 0.545& 1.4403& 4.783& 1.48E+10&\\
1.00& 2.0& 0.526& 1.2845&  3.152&  1.72E+10&\\
&&&&&&\\
1.26& 1.1& 0.943& 1.9801& 4.882& 5.12E+09 & *\\
1.26& 1.5& 0.911& 1.8484& 11.27& 5.98E+09&\\
1.26& 2.0& 0.878& 1.6508& 7.412& 6.62E+09&\\
&&&&&&\\ 	
1.60& 1.1& 1.511& 2.701& 27.65& 2.29E+09& *\\
1.60& 1.5& 1.460& 2.366& 25.74& 2.59E+09&\\ 
1.60& 2.0& 1.407& 2.116& 16.98& 2.84E+09&\\
\hline
1.30& 1.1& 0.800& 1.7710& 1.43&  4.45E+09& *\\
1.30& 1.1& 1.000& 2.0480& 4.95&  4.49E+09& *\\
1.30& 1.5& 0.800& 1.8739& 11.00& 5.55E+09& *\\
1.30& 1.5& 1.000& 1.9238& 13.09& 5.20E+09&\\
1.30& 2.0& 0.800& 1.6993& 5.989& 6.37E+09& *\\
1.30& 2.0& 1.000& 1.7154& 8.852& 5.69E+09&\\ 
&&&&&&\\
1.35& 1.1& 0.800& 1.7809& 1.272& 3.87E+09& *\\
1.35& 1.1& 1.000& 2.0794& 4.070& 3.96E+09& *\\
1.35& 1.1& 1.300& 2.2287& 12.46& 3.97E+09& *\\
1.35& 1.5& 0.800& 1.9263& 5.908& 4.91E+09& *\\
1.35& 1.5& 1.000& 1.9981& 14.01& 4.64E+09&\\
1.35& 2.0& 0.800& 1.7397& 5.010& 5.68E+09& *\\
1.35& 2.0& 1.000& 1.7822& 9.471& 5.11E+09&\\
&&&&&&\\
1.40& 1.5& 0.800& 1.9840& 5.492& 4.39E+09& *\\
1.40& 1.5& 1.000& 2.0750& 15.01& 4.14E+09&\\
1.40& 2.0& 0.800& 1.7948& 4.714& 4.96E+09& *\\                         
1.40& 2.0& 1.000& 1.8528& 10.18& 4.59E+09&\\
&&&&&&\\
1.45& 1.5& 0.800& 2.0413& 5.048& 4.00E+09& *\\
1.45& 1.5& 1.000& 2.1548& 16.09& 3.70E+09&\\ 
1.45& 1.5& 1.300& 2.1489& 19.69& 3.55E+09&\\
1.45& 2.0& 0.800& 1.8458& 4.290& 4.72E+09& *\\
1.45& 2.0& 1.000& 1.9245& 10.91& 4.11E+09&\\
1.45& 2.0& 1.300& 1.9191& 13.39& 3.84E+09&\\
&&&&&&\\
1.50& 1.5& 0.800& 2.0630& 3.753& 3.68E+09& *\\
1.50& 1.5& 1.000& 2.2350& 16.82& 3.29E+09& *\\
1.50& 1.5& 1.300& 2.2315& 21.12& 3.16E+09&\\
1.50& 2.0& 0.800& 1.8521& 2.898& 4.48E+09& *\\
1.50& 2.0& 1.000& 1.9992& 11.70& 3.67E+09&\\ 
1.50& 2.0& 1.300& 1.9938& 14.36& 3.41E+09&\\
&&&&&&\\
1.55& 1.5& 0.800& 2.0747& 2.795& 3.39E+09& *\\
1.55& 1.5& 1.000& 2.3216& 18.23& 2.94E+09& *\\
1.55& 1.5& 1.300& 2.3172& 22.67& 2.82E+09&\\
1.55& 2.0& 0.800& 1.8428& 1.979& 4.23E+09& *\\
1.55& 2.0& 1.300& 2.0704& 15.41& 3.08E+09&\\
\hline
\end{tabular}
\end{table}

In Table \ref{end-MS}, we summarize some system parameters 
when the secondaries leave the main sequence, i.e. 
the mass of the secondary, the orbital period and the age. 
We see that, when the secondaries leave the main sequence, 
many binaries are short-orbital-period systems, 
some of which are still in mass transfer. 
The distribution of periods is beyond our consideration here  
as it is related to initial orbital periods, 
but we may obtain a general ranges from previous grid calculations for
low-- and intermediate--mass binaries \cite{han00,ch02,ch03}.
For low--mass binaries with their primary masses between 1--2$M_{\odot}$, 
case B 
evolution does not result in long-orbital-period systems,
and the periods are less than 40 days.
It means that case B evolution 
may explain BSs phenomenon in 
binaries with orbital periods in this range, 
consistent with the results of Monte-Carlo simulations\cite{pol94,leo96}.
 
\section{F190}
F190 is among the five bluest stragglers in old open cluster M67.
It has been suspected of being a spectroscopic binary 
with an orbital period of about 4 days.
An orbital solution of the star and its elements were published in 1992
\cite{ml92} in which the period was confirmed to be about 4.18 days.
Tables \ref{2} and \ref{3} list some 
observations in recent years for F190 and M67, respectively.   

\begin{table}
\caption{Observational characters of F190.
The third column gives the the references.
1-Gilliland et al.(1991), 2-Milone \& Latham (1994),
3-Landsman et al.(1998), 4-Milone\& Latham (1992).}
\begin{tabular}{ccccc}
\hline
& $V$ & 10.92 - 10.95&1, 2, 3&\\
& $B-V$ &0.22 - 0.25&1, 2, 3&\\
& $P$   & 4.18284 \underline{+} 0.00015(d)&4&\\
& $e$   & 0.205 \underline{+} 0.043&4&\\
& $f(M)$& 0.0015 \underline{+} 0.0002($M_\odot$)&4&\\
& $T_{\rm eff}$& 7750(K) or 7610(K)&3&\\
& ${\rm log}g$&3.78&3&\\
\hline
\label{2}
\end{tabular}
\end{table}

\begin{table}
\caption{Observational characters of M67. 
$m-M$ is the distance modulus. R represents references.
1, Carraro et al.(1996),
2-Janes \& Smith (1984)
and 3-Fan et al. (1996).}
\begin{tabular}{ccccc}
\hline
  $Age$(Gyrs)&$m-M$ &$E(B-V)$&[Fe/H]& R\\ 
\hline
   4&9.65 & 0.025 &-0.1&1\\
   &9.48&0.056\underline{+}0.006& -0.05\underline{+}0.03&2\\
   4&9.47\underline{+}0.16&0.015--0.052&-0.10&3\\
\hline
\label{3}
\end{tabular}
\end{table}
  
Milone \& Latham \shortcite{ml92} estimated the mass of the primary (the BS)
to be in the range of 2.0 to $2.2M_\odot$ based on two extreme assumptions 
on evolutionary position and amount of light contributed by the secondary.
If we suppose that mass transfer began fairly recently, and that the turnoff
mass (of M67) was only slightly larger than $1.26M_\odot$ -- the turnoff mass 
according to VandenBerg\shortcite{van85}, 
then the initial mass of the binary could not
exceed about $2.5M_\odot$, which means that mass transfer has been quite 
efficient and very little mass was lost from the system \cite{ml92}. 
In this section, we simulate mass transfer history of this object 
based on assumption(c) as above.
  
Following constraints are considered 
when we define the ranges of our data cube:

-- the mass of the primary should be larger than the turnoff mass of M67.
Meanwhile, previous calculations \cite{han00,ch02,ch03} show that 
a mass of 1.6$M_\odot$  for the primary is too large, 
the orbital period of a binary with primary mass of 1.6$M_\odot$ 
is always larger than 4 days 
at a age longer than 3 Gyrs via case B evolution.     
The primary mass then ranges from 1.30 to 1.55$M_\odot$ 
at intervals of 0.05$M_\odot$.
We will see later that, even for case A evolution, 
$1.6M_\odot$ is an appropriate upper limit of the primary for F190. 

-- the mass ratio of the systems should not be very large in order to avoid 
the formation of a common envelope.
The chosen initial mass ratio is 1.1, 1.5 and 2.0 in our calculations.
   
-- the initial orbital period is 0.8, 1.0 and 1.3 days
for each system to ensure the onset of RLOF during HG or near to HG
(at the end of main sequence).
 
\begin{table*}
\begin{minipage}{20cm}
\caption{Parameters of the secondaries 
at orbital period of about 4.2 days.}
\begin{tabular}{ccccccccccccccc}
\hline
 $M_{\rm {1i}}$ &$P_{\rm i}$ &$q_{\rm i}$ 
  &$M_{\rm 2}$ &$X^{\rm c}_{\rm H}$ &$t$ &$M_{\rm v}$&$B-V$&
$U-B$&$X^{\rm s}_{\rm He}$&$X^{\rm s}_{\rm C}$&$X^{\rm s}_{\rm N}$&
$X^{\rm s}_{\rm O}$&$\dot{M}$ \\
($M_\odot$)&(days)&&$(M_\odot)$&&$(10^9{\rm yrs})$&&&&&&&&$(M_\odot yr^{-1})$\\
\hline
1.30& 0.80 & 1.1 & contact\\
& 1.00 &1.1& 2.016& 0.046& 4.45 & 1.317 & 0.204 &0.125&0.335&0.00176&0.00316&0.00989&8.790E-10\\
& 0.80 & 1.5& 1.814 & 0.191& 5.35 & 1.800 & 0.252&0.098&0.343&0.00142&0.00342&0.00996&4.551E-10\\
& 1.00 & 1.5& 1.779 & 0.480& 4.60 & 2.026 & 0.125&0.085&0.343&0.00151&0.00341&0.00988&6.084E-10\\
& 0.80 & 2.0& 1.661 & 0.166& 6.16 & 2.178 & 0.338&0.017&0.353&0.00108&0.00376&0.00996&2.568E-10\\
& 1.00 & 2.0& 1.632 & 0.517& 4.76 & 2.394 & 0.192&0.075&0.357&0.00116&0.00378&0.00987&3.758E-10\\
&&&&&&&&&&&\\
1.35 & 0.80 & 1.1&  contact \\
& 1.00 & 1.1& 2.086& 0.000& 3.97 & 0.936 & 0.057 &0.091&0.331&0.00175&0.00307&0.00995&9.229E-10\\
& 1.30 & 1.1& 2.025& 0.288&3.76  & 1.388 & 0.088 &0.102&0.326&0.00192&0.00289&0.00995&1.421E-09\\
& 0.80 & 1.5& 1.878& 0.122&4.80  & 1.636 & 0.261 &0.106&0.340&0.00138&0.00342&0.00998&4.694E-10\\
& 1.00 & 1.5& 1.842& 0.449&4.14  & 1.871 & 0.109 &0.088&0.339&0.00149&0.00336&0.00993&6.175E-10\\
& 0.80 & 2.0& 1.719& 0.073& 5.59 &2.022 &0.355 &0.010&0.349&0.00105&0.00376&0.00997&2.810E-10\\
& 1.00 & 2.0& 1.690& 0.473& 4.36 &2.225 &0.175 &0.082&0.352&0.00113&0.00377&0.00988&3.872E-10\\
&&&&&&&&&&&\\
1.40 & 0.80 & 1.5& 1.944& 0.101& 4.31 & 1.469 & 0.247 &0.115&0.338&0.00129&0.00355&0.00992&5.065E-10\\
& 1.00 & 1.5& 1.906& 0.418& 3.73 & 1.719 & 0.092 &0.088&0.336&0.00142&0.00348&0.00986&6.571E-10\\
& 0.80 & 2.0& 1.781& 0.042& 4.92 & 1.843 & 0.329 &0.034&0.346&0.00096&0.00389&0.00992&3.037E-10\\
& 1.00 & 2.0& 1.750& 0.451& 3.98 & 2.057 & 0.152 &0.089&0.348&0.00107&0.00386&0.00983&4.166E-10\\
&&&&&&&&&&&\\
1.45& 0.80 & 1.5& 2.012 & 0.073& 3.95 & 1.316 & 0.222 &0.124&0.339&0.00122&0.00364&0.00989&5.684E-10\\
& 1.00 & 1.5& 1.973& 0.400& 3.34 & 1.581 & 0.075 &0.080&0.335&0.00139&0.00346&0.00989&7.164E-10\\
& 1.30 & 1.5& 1.920& 0.572& 2.95 & 1.850 & 0.048 &0.041&0.334&0.00154&0.00337&0.00984&1.174E-09\\
& 0.80 & 2.0& 1.843& 0.001& 4.71 & 1.413 & 0.151 &0.130&0.349&0.00087&0.00403&0.00985&3.434E-10\\
& 1.00 & 2.0& 1.812& 0.427& 3.61 & 1.909 & 0.129 &0.092&0.346&0.00104&0.00388&0.00983&4.438E-10\\
& 1.30 & 2.0& 1.770& 0.621& 2.96 & 2.190 & 0.096 &0.069&0.348&0.00115&0.00385&0.00976&7.801E-10\\
&&&&&&&&&&&\\
1.50& 0.80 & 1.5& 2.083& 0.000& 3.71 &0.794 & 0.123& 0.794&0.339&0.00114&0.00376&0.00983&6.135E-10\\
& 1.00 & 1.5& 2.042&0.389&2.98 &1.452&0.055&0.065&0.334&0.00133&0.00385&0.00984&7.928E-10\\ 
& 1.30 & 1.5& 1.987&0.564&2.62 &1.730 &0.029&0.019&0.333&0.00148&0.00347&0.00980&1.318E-09\\
& 0.80 & 2.0& 1.908&0.000&4.62 &1.691 &0.978& 0.712&0.307&0.00204&0.00629&0.00996&3.959E-10\\
& 1.00 & 2.0& 1.876&0.409&3.24 &1.768 &0.107 &0.092&0.345&0.00098&0.00398&0.00977&4.933E-10\\
& 1.30 & 2.0& 1.832&0.615&2.64 & 2.055& 0.071&0.058&0.346&0.00110&0.00395&0.00970&8.653E-10\\ 
&&&&&&&&&&&\\
1.55 & 0.80 & 1.5& contact\\
& 1.00 & 1.5& 2.114& 0.415&2.63 &1.362 &0.025 &0.028&0.336&0.00123&0.00376&0.00974&9.226E-10\\
& 1.30 & 1.5& 2.057& 0.573& 2.34 &1.613 &0.010&-0.002&0.334&0.00142&0.00366&0.00973&1.486E-09\\
& 0.80 & 2.0& contact\\
& 1.00 & 2.0& contact\\
& 1.30 & 2.0& 1.896& 0.611&2.37 &1.938 &0.047 &0.036&0.348&0.00104&0.00412&0.00961&9.924E-10\\
\hline
\label{5}
\end{tabular}
\end{minipage}
\end{table*}

Characteristics for all of the binaries during RLOF are listed 
in Table \ref{6}.    
Some parameters of the secondaries are shown in Table \ref{5} 
when the orbital period is about 4.2 days.
Visual magnitude in V band $M_{\rm v}$ is calculated by 
\begin{eqnarray}
M_{\rm v}=M_{\rm bol}-BC,& M_{\rm bol}= 4.75-2.5\times {\rm log}(L/L_\odot),
\end{eqnarray} 
$BC$, $B-V$ and $U-B$ are obtained by linear interpolation 
from BaSel-2.0 model \cite{basel97,basel98} 
via temperature $T$ and ${\rm log}g$.
Although 
the metallicity of M67 is a little different from solar (Table \ref{3}),
the difference is small.
Meanwhile some other studies concur that 
the metallicity of M67 is virtually indistinguishable from solar, 
i.e. $\rm [Fe/H]=-0.09\underline{+}0.07$ 
according to Friel \& Janes\shortcite{fj93},
$\rm [Fe/H]=-0.04\underline{+}0.12$ 
according to Hobbs \& Thorburn \shortcite{ht91}.
Therefore we choose the table of $\rm [Fe/H]=0.0$ when we interpolate, 
consistent with the model calculations.  

In Table \ref{5}, we see that the surface abundances 
of He, C, N, O are obviously abnormal
(the normal values are 0.28000, 0.00352, 0.00106 and 0.01004, respectively). 
Meanwhile, all the models are in slow mass transfer with a 
rate $\sim 10^{-10}M_\odot {\rm yr^{-1}}$, which is consistent to
the argument of Milone \& Latham \shortcite{ml92} that a final slow stage of 
mass transfer is still underway for the F190.
We also see that, when ($P_{\rm i},q_{\rm i}$) is constant, 
the gainer is younger at the period of 4.2 days 
with the increase of primary's mass 
until the system becomes a contact binary finally. 
As $M_{\rm 1i}>1.5M_\odot$, 
the system is either contact or younger than 3 Gyrs when 
the orbital period is about 4.2 days. 
The value of $1.6M_\odot$ is therefore a `safe' upper limit 
of the primary mass as we study the mass-transfer history of F190,
including case A evolution. 
 
\begin{figure}
\centerline{\psfig{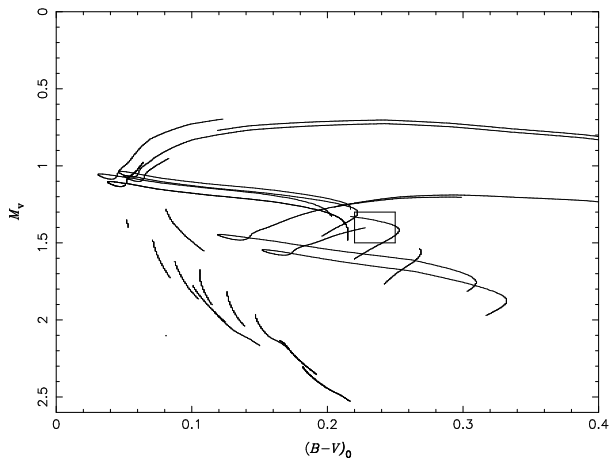}}
\caption{Evolutionary tracks of the secondaries on color-magnitude diagram
when the system's orbital period is between 3.2 days and 5.2 days
and the age changes from $3\times10^9$ to $5\times10^9$ yrs. 
Observational range of F190 is determined by 
a distance modulus $m-M=9.45-9.65$ with $V=10.95$, $B-V=0.22-0.25$.
Three lines go across the range of F190.
Parameters for the three lines (from the bottom right to the top left) are 
$(M_{\rm1i},q_{\rm i},P_{\rm i})=(1.40,1.5,0.8),(1.40,2.0,0.8),(1.45,1.5,080)$, respectively.}
\label{f1902}
\end{figure}

To find out the appropriate models for F190, 
we show the location of F190 in the model grid in figure \ref{f1902} 
with some constraints included, i.e.  
the system's orbital period ($3.2-5.2$ days)
and the age of the models ($3-5\times10^9$ yrs)
are around 4.2 days and 4 Gyrs, respectively.
The flexible ranges of orbital period and age are chosen here 
as there are errors between theory models and observations, e.g.,
due to the coarseness of the grid the models cannot be expected to fit 
the observations exactly.
The position range of F190 by observations is also shown in the figure. 
The distance modulus $m-M$ of M67 ranges from 9.45 to 9.65 while $V=10.95$,
$B-V$ ranges from 0.22 to 0.25.
Three evolutionary tracks under these constraints 
pass through the area of F190 in the figure.
The initial parameters for the three models are 
$(M_{\rm1i},q_{\rm i},P_{\rm i})=(1.40,1.5,0.8),(1.40,2.0,0.8),(1.45,1.5,0.8)$ 
from the bottom right to the top left 
in the area of F190, respectively.  
However, 
the secondary in the model with 
$(M_{\rm1i},q_{\rm i},P_{\rm i})=(1.40,2.0,0.8)$
has already left the main sequence and cannot be considered as a BS.
The primary's mass of appropriate models for F190 is then located in 
the range of 1.40 to 1.45$M_{\odot}$ with $(q_{\rm i},P_{\rm i})=(1.5,0.8)$.
The best-fitting model 
is a binary with $(M_{\rm 1i},q_{\rm i},P_{\rm i})=(1.45,1.5,0.8)$. 
Parameters of the system at the orbital period about of 4.2 days are 
 $M_{\rm 2}=2.0M_\odot$, $M_{\rm v}=1.316$, $B-V=0.222$ and 
the age $t=3.95\times 10^9$ yrs, which fit the observations well.

\begin{figure}
\centerline{\psfig{figure=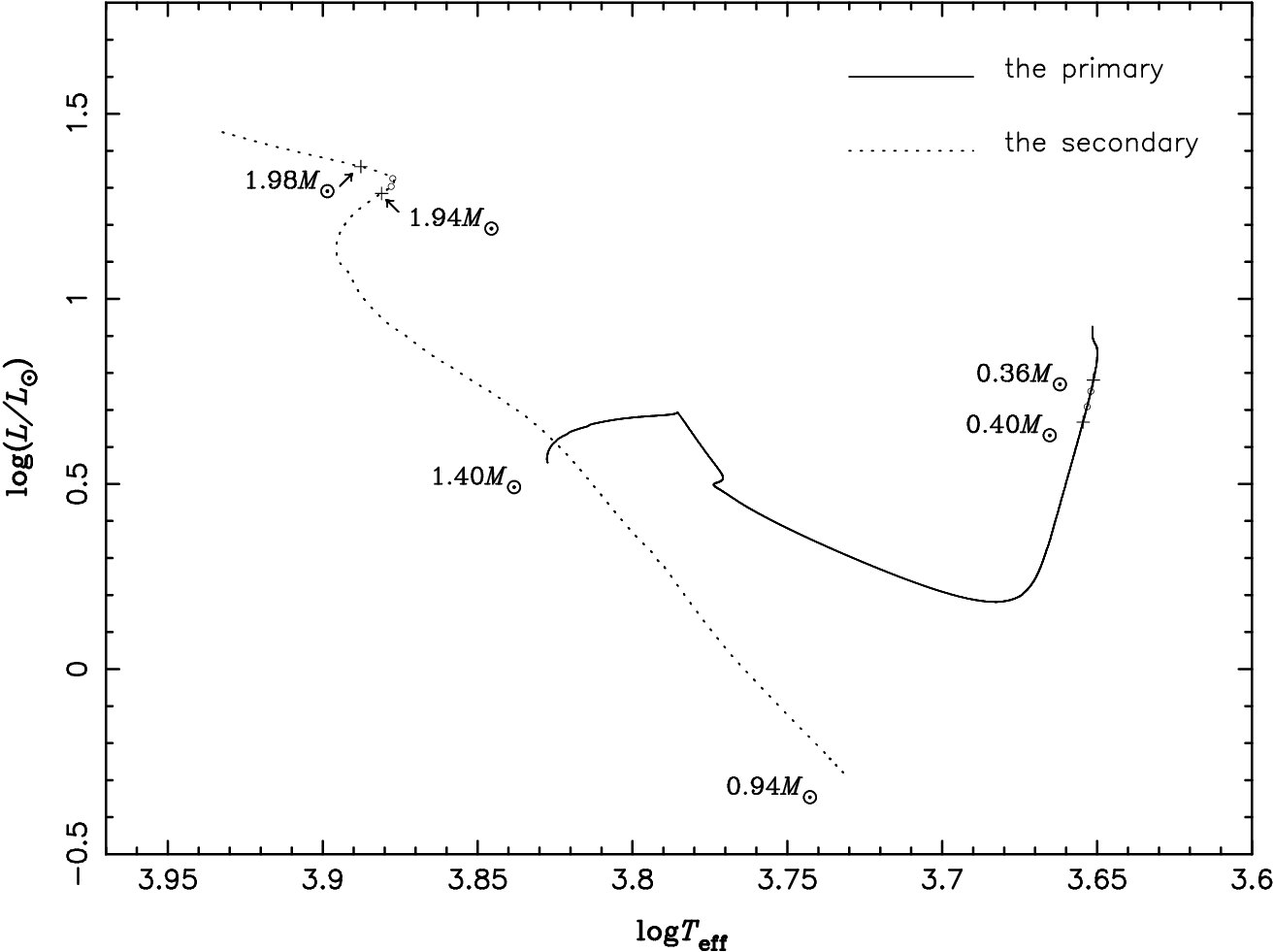,width=8cm,angle=270}}
\caption{Evolutionary tracks of both components for a binary $1.40+0.93M_\odot$
with an initial orbital period 0.80 days. 
The part between two pluses, except the part between two circles, 
represent the secondary stays in the region of F190 by observation.}
\label{f1903}
\end{figure}

\begin{figure}
\centerline{\psfig{figure=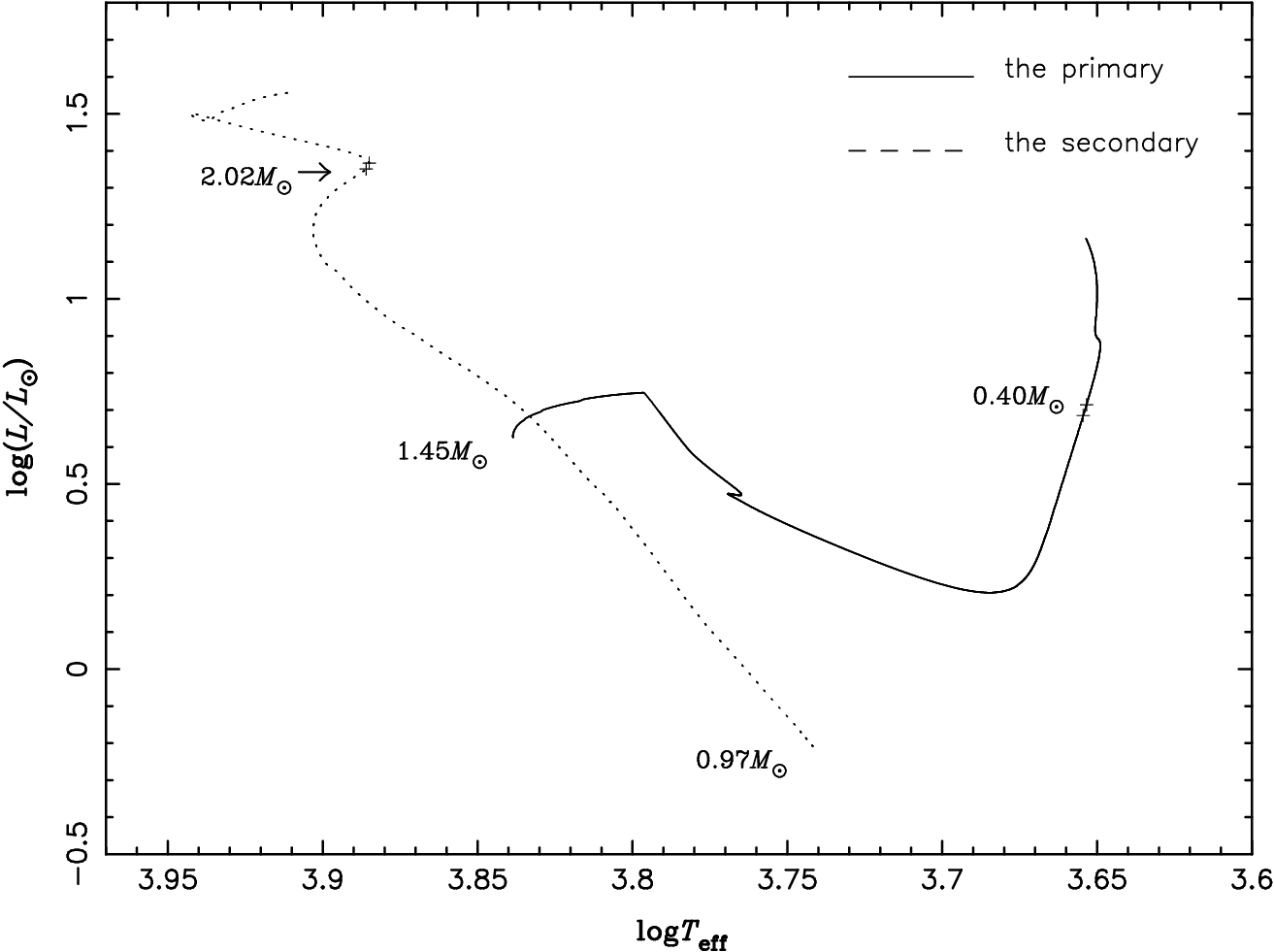,width=8cm,angle=270}}
\caption{Evolutionary tracks of both components for a binary $1.45+0.97M_\odot$
with an initial orbital period 0.80 days. 
The part between two pluses represent the secondary stays in the region 
of F190 by observation.}
\label{f1904}
\end{figure}

Figures \ref{f1903} and \ref{f1904} show the evolutionary tracks of binaries 
$(M_{\rm1i},q_{\rm i},P_{\rm i})=(1.40,1.5,0.8)$ 
and (1.45,1.5,0.8), respectively. 
The locations of both components
as the secondaries is in the observational region of F190
are ranged out by pluses in the two figures.
We see that the surface effective temperature of the secondaries is  
in $3.88<{\rm log}T_{\rm eff}<3.89$ 
if $M_{\rm 2}$ is located in the observational region of F190, 
very close to the value given by Landsman et al. \shortcite{land98}
(${\rm log}T_{\rm eff}=3.889$ and 3.881 for $T_{\rm eff}=7750$ K and 7610 K, 
respectively). 
However, the surface gravity in the two models (${\rm log}g$ 
in the region 3.87-3.91) is a little bigger than the 
value given in Table \ref{2}.  
Both the two systems begin RLOF near the end of the main sequence
(see also Table \ref{6}), 
indicating that case A is a more likely channel to form F190, 
different from what has long been suspected that
case B mass transfer leads the formation of the object \cite{ml92,leo96}.
The property of the companion ($M_{\rm 1}$) is worth a mention.    
In figure 13, $M_{\rm v}=3.65$ and 3.41 for $M_{\rm 1}$ as
its mass equal to 0.40 and 0.36$M_\odot$, respectively.
In figure 14, $M_{\rm v}$ of the primary is about 3.55 
as the secondary in the observational area of F190.
This result means that 
it is hard to detect the signal from the companion by spectroscopic 
except for a very high ratio of signal to noise (e.g. $S/N \geq 500$).
Meanwhile, IUE spectra \cite{land98}
show no evidence for a hot subluminous companion for F190, 
indicating that the companion is likely a cool star,
which is consistent with the position 
shown in the figures \ref{f1903} and \ref{f1904} .            
                                           
Orbital solutions of F190 \cite{ml92} showed that
the object is an eccentric binary.
Mass transfer cannot explain the phenomenon  
since tidal effects should have circularized the orbit of a binary 
with such a short orbital period.
Milone \& Latham \shortcite{ml92} mentioned that 
the eccentric orbit might be 
an artifact caused by some sort of line asymmetries in orbital solutions, 
or a resonant phenomena in an accretion disk,  
or a consequence of modulation by a distant third star in a wide orbit.
Hurley et al. \shortcite{hur01} obtained 
two BSs in short-period eccentric orbits via a collision 
within a triple system, 
and suggested that F190 might be produced in this way. 
Sandquist et al. \shortcite{san03} reported that 
BS S1082 in M67 is a possible triple 
(see also van den Berg et al. \shortcite{van01}). 
Meanwhile, orbital evolution in triples shows that 
a third star may indeed induce eccentricity 
in the orbit of a close binary \cite{ppe01}.
If the eccentricity is real and induced by a triple companion 
or an encounter with a third body, then the orbital period is modified 
after RLOF. 
In this case the current orbital parameters do not result purely from
RLOF, and our procedure of fitting the period will not give the 
correct initial parameters.
          
\section{Discussions and Conclusions}
In the investigation, 
we studied the structure and evolution of the mass gainers 
on the main sequence for low--mass binaries under 
different assumptions (see section 4). 
We see that  
the decrease of hydrogen mass fraction on the surface (assumption (b)) 
makes the gainer  
bluer and smaller than the ones with original surface composition 
(assumption (a)) 
while thermohaline mixing (assumption (c)) lessens the effect slightly.
If thermohaline mixing were to act instantaneously (assumption (d)), 
the effect would be lessened more.
However the effect is small and probably cannot be observed. 
Meanwhile, 
thermohaline mixing (either instantaneous or not) 
causes composition redistribution in the envelopes, 
but not in the outer most regions during RLOF
because of continuous accretion.
Therefore obvious CNO abundance abnormalities, 
originating from matter that was originally in the inner regions of the
primary, exist on the surfaces of the secondary 
during RLOF under assumptions (b), (c) and (d).
After RLOF, CNO abundance abnormalities are greatly weakened
under assumptions (c) and (d)
by thermohaline mixing taking effect on the surface abundances,
but element contamination (from the primary) will not disappear.  
We therefore prefer to a time-dependent way 
rather than an instantaneous process in treating thermohaline mixing 
when we study the BSs formed by mass transfer.  

Some secondaries are still near the end of RLOF at the age larger 
than that of M67 in our simulations. 
The orbital period after RLOF in our calculations
and previous results for low--mass binaries 
ranges from several days to tens of days.
It means that case A and case B mass transfer 
may produce BSs in short- or relatively short-orbital-periods  
binaries (including Algol systems), 
i.e. the orbital period less than forty days.  
This result consistent with the Monte-Carlo simulation by some authors
\cite{pol94,leon96}.
Further constraints to the range of orbital period of BSs 
formed via this channel need binary population synthesis.  

As we simulate BSs formed via binary mass transfer,
some physical constraints should be considered 
to limit initial models.
At first, 
the system should avoid common envelope -- 
a more complex case and we will study it in the future. 
It means that the initial mass ratio should not be very large 
and the primary has a radiative envelope when RLOF starts.
Secondly,
the primary cannot be much more massive than the turn-off mass of the cluster,
or it evolves  very fast, 
the secondary then increases to a high mass in a very short time
and evolves rapidly after a short self-adjustment.
As a consequence, the secondary should have left the main sequence 
at the age of the host cluster, then cannot be recognized as a BS.
Combining these considerations with observations of F190 and M67, 
we ranged the initial parameters of the object F190 
by complete grid simulations.
We find that 
the primary's mass in appropriate models for F190 is located in 
the range of 1.40 to 1.45$M_{\odot}$ with $(q_{\rm i},P_{\rm i})=(1.5,0.8)$
and that case A evolution is a more likely evolutionary channel than case B
to form this object. 
Evolutionary locations of both components for two appropriate models 
are presented in the paper to help determine the companion of F190.
Our simulation indicates that F190 is still in slow mass transfer phase 
and exists obvious CNO abundance abnormalities. 
The abnormalities might be observed later. 

As there are very few BSs in short-orbital-period systems observationally,
binary merger (case A) and case C mass transfer may be the main sources of BSs
among binary interactions. 
However binary merger is a very complex process during which
much physics is not very clear yet.
An initial model of a merger can just be constructed 
based on relatively clear physical phases and some assumptions, 
i.e. the degree of matter mixing.
For case C mass transfer, finding out the region of initial mass ratio to 
avoid dynamical instability is important.
We will study both of the two cases in our following work.

\section*{acknowledgments}
We thank Prof. B. Shustov for his kind help.
We also thank Prof. S. M. Andervisky, Dr. Ph. Podsiadlowski, Dr. F. Zhang
and Sh. Gu 
for their helpful suggestions and discussions on the subject. 
We are grateful to the referee for his useful and inspirational suggestions
which help us improve the paper greatly.
This work is supported 
by the Chinese National Science Foundation (Grant No. 19925312), 
the Chinese Academy of Sciences(No. KJCX2-SW-T06), 
and the 973 Scheme (NKBRSF No. G19990754).

\appendix

\section{RLOF for all the binaries}

Table A1 lists all the binaries simulating F190 in Sect.5: 
\begin{description}
\item a: the onset of RLOF
\item b: the minimum luminosity during RLOF (i.e., just reaching the RGB)
\item c: at the end of RLOF 
         (RLOF may have several episodes, there is no RLOF any more after c)
\item d: at the end of calculation 
\end{description}
We usually list stellar parameters at a, b, c and d. However,
the code breaks down
when RLOF is unstable. In that case, we only list parameters at a 
if RLOF is unstable at onset, or at a and b 
if RLOF is stable at the onset but becomes unstable after primary reaching RGB. 
\begin{table*}
\begin{minipage}{12cm}
\caption{RLOF of all the binaries simulated for a BS of F190.}
\begin{tabular}{ccccccccccc}
\hline
 & $\scriptstyle t_1$ & $\scriptstyle M_1$ &
   $\scriptstyle \dot{M}_1$ &
   $\scriptstyle X_{\rm H}^{\rm c}$ & $\scriptstyle X_{\rm H}^{\rm s}$ & 
   $\scriptstyle q$ & $\scriptstyle M_2$ &
   $\scriptstyle P$\\
 & $\scriptstyle ({\rm yr})$ & $\scriptstyle (M_\odot)$ &
   $\scriptstyle (M_\odot {\rm yr}^{-1})$ & & & & 
   $\scriptstyle (M_\odot)$ & 
   $\scriptstyle ({\rm days})$  \\
\hline
a&$ 3.334  \times 10^{  9}$ &1.303&$ 0.000  $                &0.000&0.700&1.096&1.189& 0.8000      \\
b&$ 4.268  \times 10^{  9}$ &0.866&$-8.626  \times 10^{-10}$ &0.000&0.688&0.533&1.625& 1.064     \\
c&$ 5.224  \times 10^{  9}$ &0.256&$-3.507  \times 10^{-12}$ &0.000&0.644&0.115&2.236& 15.84      \\
d&$ 5.253  \times 10^{  9}$ &0.256&$ 0.000  $                &0.000&0.644&0.115&2.236& 15.84      \\
&&&&&&&&\\
a&$ 3.717  \times 10^{  9}$ &1.303&$ 0.000  $                &0.000&0.700&1.096&1.189& 1.000       \\
b&$ 4.071  \times 10^{  9}$ &0.920&$-2.161  \times 10^{ -9}$ &0.000&0.687&0.585&1.572& 1.229       \\
c&$ 4.770  \times 10^{  9}$ &0.261&$-1.113  \times 10^{-12}$ &0.000&0.646&0.117&2.231& 18.79      \\
d&$ 4.795  \times 10^{  9}$ &0.261&$ 0.000  $                &0.000&0.646&0.117&2.231& 18.79      \\
&&&&&&&&\\
a&$ 4.034  \times 10^{  9}$ &1.303&$ 0.000  $                &0.000&0.699&1.096&1.189& 1.300      \\
b&$ 4.039  \times 10^{  9}$ &1.251&$-5.337  \times 10^{-10}$ &0.000&0.699&1.009&1.240& 1.292      \\
&&&&&&&&\\
a&$ 3.356  \times 10^{  9}$ &1.303&$ 0.000  $                &0.000&0.700&1.496&0.871& 0.8000       \\
b&$ 4.428  \times 10^{  9}$ &0.761&$-5.598  \times 10^{-10}$ &0.000&0.682&0.539&1.413& 0.9406   \\
c&$ 5.700  \times 10^{  9}$ &0.246&$-2.413  \times 10^{-12}$ &0.000&0.637&0.127&1.928& 11.00      \\
d&$ 5.754  \times 10^{  9}$ &0.246&$ 0.000  $                &0.000&0.637&0.127&1.928& 11.00      \\
&&&&&&&&\\
a&$ 3.768  \times 10^{  9}$ &1.303&$ 0.000  $                &0.000&0.700&1.496&0.871& 1.000        \\
b&$ 4.050  \times 10^{  9}$ &0.817&$-1.514  \times 10^{ -9}$ &0.000&0.684&0.602&1.357& 1.073      \\
c&$ 4.939  \times 10^{  9}$ &0.250&$-1.120  \times 10^{-12}$ &0.000&0.637&0.130&1.924& 13.09      \\
d&$ 4.976  \times 10^{  9}$ &0.250&$ 0.000  $                &0.000&0.637&0.130&1.924& 13.09      \\
&&&&&&&&\\
a&$ 4.048  \times 10^{  9}$ &1.303&$ 0.000  $                &0.000&0.699&1.496&0.871& 1.300      \\
&&&&&&&&\\
a&$ 3.435  \times 10^{  9}$ &1.303&$ 0.000  $                &0.000&0.700&2.000&0.652& 0.8000      \\
b&$ 4.607  \times 10^{  9}$ &0.667&$-2.547  \times 10^{-10}$ &0.000&0.678&0.518&1.288& 0.7734    \\
c&$ 6.488  \times 10^{  9}$ &0.235&$ 0.000  $                &0.000&0.627&0.137&1.720& 7.410      \\
d&$ 2.859  \times 10^{ 10}$ &0.235&$ 0.000  $                &0.000&0.627&0.137&1.720& 7.410      \\
&&&&&&&&&\\
a&$ 3.814  \times 10^{  9}$ &1.303&$ 0.000  $                &0.000&0.700&2.000&0.652& 1.000      \\
b&$ 3.982  \times 10^{  9}$ &0.700&$-9.078  \times 10^{-10}$ &0.000&0.678&0.558&1.255& 0.9034    \\
c&$ 5.107  \times 10^{  9}$ &0.239&$-1.569  \times 10^{-12}$ &0.000&0.623&0.139&1.715& 8.852      \\
d&$ 2.636  \times 10^{ 10}$ &0.239&$ 0.000  $                &0.000&0.623&0.139&1.715& 8.852      \\
&&&&&&&&&\\
a&$ 4.065  \times 10^{  9}$ &1.303&$ 0.000  $                &0.000&0.698&2.000&0.652& 1.300      \\
&&&&&&&&&\\
a&$ 2.959  \times 10^{  9}$ &1.349&$ 0.000  $                &0.017&0.700&1.097&1.230& 0.8000   \\
b&$ 3.776  \times 10^{  9}$ &0.883&$-9.677  \times 10^{-10}$ &0.000&0.688&0.520&1.696& 1.089       \\
c&$ 4.708  \times 10^{  9}$ &0.258&$ 0.000  $                &0.000&0.647&0.111&2.321& 16.97      \\
d&$ 4.735  \times 10^{  9}$ &0.258&$ 0.000  $                &0.000&0.647&0.111&2.321& 16.97      \\
&&&&&&&&\\
a&$ 3.193  \times 10^{  9}$ &1.349&$ 0.000  $                &0.000&0.700&1.097&1.230& 1.000        \\
b&$ 3.597  \times 10^{  9}$ &0.940&$-2.156  \times 10^{ -9}$ &0.000&0.687&0.574&1.639& 1.249    \\
c&$ 4.284  \times 10^{  9}$ &0.263&$-3.791  \times 10^{-12}$ &0.000&0.649&0.114&2.316& 20.13      \\
d&$ 4.308  \times 10^{  9}$ &0.263&$ 0.000  $                &0.000&0.649&0.114&2.316& 20.13      \\
&&&&&&&&\\
a&$ 3.531  \times 10^{  9}$ &1.349&$ 0.000  $                &0.000&0.700&1.097&1.230& 1.300        \\
b&$ 3.568  \times 10^{  9}$ &1.009&$-6.358  \times 10^{ -9}$ &0.000&0.686&0.642&1.571& 1.495   \\
c&$ 4.045  \times 10^{  9}$ &0.270&$-6.628  \times 10^{-12}$ &0.000&0.653&0.117&2.310& 24.61      \\
d&$ 4.066  \times 10^{  9}$ &0.270&$ 0.000  $                &0.000&0.653&0.117&2.310& 24.61      \\
&&&&&&&&\\
a&$ 2.989  \times 10^{  9}$ &1.349&$ 0.000  $                &0.003&0.700&1.496&0.902& 0.8000      \\
b&$ 3.897  \times 10^{  9}$ &0.785&$-5.940  \times 10^{-10}$ &0.000&0.684&0.535&1.466& 0.9457    \\
c&$ 5.148  \times 10^{  9}$ &0.248&$-1.002  \times 10^{-12}$ &0.000&0.640&0.124&2.003& 11.78      \\
d&$ 5.190  \times 10^{  9}$ &0.248&$ 0.000  $                &0.000&0.640&0.124&2.003& 11.78      \\
&&&&&&&&&\\
a&$ 3.238  \times 10^{  9}$ &1.349&$ 0.000  $                &0.000&0.700&1.496&0.902& 1.000       \\
b&$ 3.596  \times 10^{  9}$ &0.826&$-1.424  \times 10^{ -9}$ &0.000&0.683&0.580&1.424& 1.103    \\
c&$ 4.480  \times 10^{  9}$ &0.252&$-3.474  \times 10^{-12}$ &0.000&0.641&0.126&1.998& 14.01      \\
d&$ 4.514  \times 10^{  9}$ &0.252&$ 0.000  $                &0.000&0.641&0.126&1.998& 14.01      \\
\hline
\multicolumn{9}{r}{continued to next paper}
\label{6}
\end{tabular}
\end{minipage}
\end{table*}

\setcounter{table}{0}
\begin{table*}
\begin{minipage}{12cm}
\caption{-continued}
\begin{tabular}{ccccccccccc}
\hline
 & $\scriptstyle t_1$ & $\scriptstyle M_1$ &
   $\scriptstyle \dot{M}_1$ &
   $\scriptstyle X_{\rm H}^{\rm c}$ & $\scriptstyle X_{\rm H}^{\rm s}$ & 
   $\scriptstyle q$ & $\scriptstyle M_2$ &
   $\scriptstyle P$ \\
 & $\scriptstyle ({\rm yr})$ & $\scriptstyle (M_\odot)$ &
   $\scriptstyle (M_\odot {\rm yr}^{-1})$ & & & & 
   $\scriptstyle (M_\odot)$ & 
   $\scriptstyle ({\rm days})$ \\
\hline
a&$ 3.547  \times 10^{  9}$ &1.349&$ 0.000  $                &0.000&0.699&1.496&0.902& 1.300&   \\
&&&&&&&&\\
a&$ 2.994  \times 10^{  9}$ &1.349&$ 0.000  $                &0.000&0.700&2.000&0.675& 0.8000     \\
b&$ 4.113  \times 10^{  9}$ &0.680&$-2.742  \times 10^{-10}$ &0.000&0.678&0.506&1.343& 0.7901   \\
c&$ 5.933  \times 10^{  9}$ &0.237&$-1.151  \times 10^{-12}$ &0.000&0.631&0.133&1.786& 7.934      \\
d&$ 2.782  \times 10^{ 10}$ &0.237&$ 0.000  $                &0.000&0.631&0.133&1.786& 7.934      \\
&&&&&&&&&\\
a&$ 3.285  \times 10^{  9}$ &1.349&$ 0.000  $                &0.000&0.700&2.000&0.675& 1.000      \\
b&$ 3.534  \times 10^{  9}$ &0.721&$-7.896  \times 10^{-10}$ &0.000&0.678&0.553&1.303& 0.9099    \\
c&$ 4.709  \times 10^{  9}$ &0.241&$-1.727  \times 10^{-12}$ &0.000&0.628&0.135&1.782& 9.471      \\
d&$ 2.583  \times 10^{ 10}$ &0.241&$ 0.000  $                &0.000&0.628&0.135&1.782& 9.471      \\
&&&&&&&&&\\
a&$ 3.563  \times 10^{  9}$ &1.349&$ 0.000  $                &0.000&0.699&2.000&0.675& 1.300      \\
&&&&&&&&&\\
a&$ 2.551  \times 10^{  9}$ &1.396&$ 0.000  $                &0.060&0.700&1.496&0.933& 0.8000     \\
b&$ 3.485  \times 10^{  9}$ &0.795&$-6.560  \times 10^{-10}$ &0.000&0.684&0.518&1.534& 0.9743    \\
c&$ 4.656  \times 10^{  9}$ &0.250&$-5.982  \times 10^{-12}$ &0.000&0.642&0.120&2.080& 12.63      \\
d&$ 4.694  \times 10^{  9}$ &0.250&$ 0.000  $                &0.000&0.642&0.120&2.080& 12.63      \\
&&&&&&&&\\
a&$ 2.789  \times 10^{  9}$ &1.396&$ 0.000  $                &0.000&0.700&1.496&0.933& 1.000      \\
b&$ 3.198  \times 10^{  9}$ &0.851&$-1.453  \times 10^{ -9}$ &0.000&0.684&0.575&1.479& 1.111     \\
c&$ 4.066  \times 10^{  9}$ &0.255&$-4.557  \times 10^{-12}$ &0.000&0.644&0.123&2.075& 15.01      \\
d&$ 4.097  \times 10^{  9}$ &0.255&$ 0.000  $                &0.000&0.644&0.123&2.075& 15.01      \\
&&&&&&&&\\
a&$ 3.103  \times 10^{  9}$ &1.396&$ 0.000  $                &0.000&0.700&1.496&0.933& 1.300      \\
&&&&&&&&\\
a&$ 2.649  \times 10^{  9}$ &1.396&$ 0.000  $                &0.014&0.700&1.995&0.700& 0.8000     \\
b&$ 3.584  \times 10^{  9}$ &0.693&$-3.196  \times 10^{-10}$ &0.000&0.679&0.494&1.403& 0.8124     \\
c&$ 5.262  \times 10^{  9}$ &0.239&$ 0.000  $                &0.000&0.633&0.129&1.857& 8.535      \\
d&$ 2.689  \times 10^{ 10}$ &0.239&$ 0.000  $                &0.000&0.633&0.129&1.857& 8.535      \\
&&&&&&&&\\
a&$ 2.834  \times 10^{  9}$ &1.396&$ 0.000  $                &0.000&0.700&1.995&0.700& 1.000      \\
b&$ 3.157  \times 10^{  9}$ &0.740&$-7.591  \times 10^{-10}$ &0.000&0.680&0.546&1.356& 0.9228     \\
c&$ 4.336  \times 10^{  9}$ &0.243&$-1.247  \times 10^{-12}$ &0.000&0.632&0.131&1.853& 10.18      \\
d&$ 2.523  \times 10^{ 10}$ &0.243&$ 0.000  $                &0.000&0.632&0.131&1.853& 10.18      \\
&&&&&&&&\\
a&$ 3.122  \times 10^{  9}$ &1.396&$ 0.000  $                &0.000&0.699&1.995&0.700& 1.300      \\
&&&&&&&&\\
a&$ 2.068  \times 10^{  9}$ &1.445&$ 0.000  $                &0.174&0.700&1.496&0.966& 0.8000    \\
b&$ 3.255  \times 10^{  9}$ &0.813&$-8.057  \times 10^{-10}$ &0.000&0.685&0.508&1.599& 0.9928    \\
c&$ 4.300  \times 10^{  9}$ &0.252&$-3.583  \times 10^{-12}$ &0.000&0.641&0.117&2.160& 13.57      \\
d&$ 4.335  \times 10^{  9}$ &0.252&$ 0.000  $                &0.000&0.641&0.117&2.160& 13.57      \\
&&&&&&&&\\
a&$ 2.449  \times 10^{  9}$ &1.445&$ 0.000  $                &0.000&0.700&1.496&0.966& 1.000     \\
b&$ 2.864  \times 10^{  9}$ &0.865&$-1.578  \times 10^{ -9}$ &0.000&0.684&0.559&1.546& 1.137      \\
c&$ 3.679  \times 10^{  9}$ &0.257&$-1.796  \times 10^{-12}$ &0.000&0.645&0.119&2.155& 16.09      \\
d&$ 3.707  \times 10^{  9}$ &0.257&$ 0.000  $                &0.000&0.645&0.119&2.155& 16.09      \\
&&&&&&&&\\
a&$ 2.662  \times 10^{  9}$ &1.445&$ 0.000  $                &0.000&0.700&1.496&0.966& 1.300      \\
b&$ 2.759  \times 10^{  9}$ &0.918&$-6.128  \times 10^{ -9}$ &0.000&0.684&0.615&1.493& 1.373    \\
c&$ 3.254  \times 10^{  9}$ &0.263&$-8.057  \times 10^{-13}$ &0.000&0.646&0.122&2.149& 19.69      \\
d&$ 3.278  \times 10^{  9}$ &0.263&$ 0.000  $                &0.000&0.646&0.122&2.149& 19.69      \\
&&&&&&&&\\
a&$ 2.174  \times 10^{  9}$ &1.445&$ 0.000  $                &0.130&0.700&1.995&0.724& 0.8000     \\
b&$ 3.640  \times 10^{  9}$ &0.707&$-4.192  \times 10^{-10}$ &0.000&0.680&0.483&1.463& 0.8300     \\
c&$ 5.065  \times 10^{  9}$ &0.241&$ 0.000  $                &0.000&0.631&0.125&1.929& 9.173      \\
d&$ 2.634  \times 10^{ 10}$ &0.241&$ 0.000  $                &0.000&0.631&0.125&1.929& 9.173      \\
&&&&&&&&\\
a&$ 2.480  \times 10^{  9}$ &1.445&$ 0.000  $                &0.000&0.700&1.995&0.724& 1.000     \\
b&$ 2.848  \times 10^{  9}$ &0.752&$-8.036  \times 10^{-10}$ &0.000&0.680&0.531&1.417& 0.9464      \\
c&$ 3.960  \times 10^{  9}$ &0.245&$-6.654  \times 10^{-13}$ &0.000&0.634&0.127&1.925& 10.91      \\
d&$ 4.015  \times 10^{  9}$ &0.245&$ 0.000  $                &0.000&0.634&0.127&1.925& 10.91      \\
\hline
\multicolumn{9}{r}{continued to next paper}
\label{7}
\end{tabular}
\end{minipage}
\end{table*}

\setcounter{table}{0}
\begin{table*}
\begin{minipage}{12cm}
\caption{--continued.}
\begin{tabular}{ccccccccccc}
\hline
 & $\scriptstyle t_1$ & $\scriptstyle M_1$ &
   $\scriptstyle \dot{M}_1$ &
   $\scriptstyle X_{\rm H}^{\rm c}$ & $\scriptstyle X_{\rm H}^{\rm s}$ & 
   $\scriptstyle q$ & $\scriptstyle M_2$ &
   $\scriptstyle P$\\
 & $\scriptstyle ({\rm yr})$ & $\scriptstyle (M_\odot)$ &
   $\scriptstyle (M_\odot {\rm yr}^{-1})$ & & & & 
   $\scriptstyle (M_\odot)$ & 
   $\scriptstyle ({\rm days})$ \\
\hline
a&$ 2.694  \times 10^{  9}$ &1.445&$ 0.000  $                &0.000&0.700&1.995&0.724& 1.300      \\
b&$ 2.732  \times 10^{  9}$ &0.795&$-5.005  \times 10^{ -9}$ &0.000&0.681&0.578&1.375& 1.143    \\
c&$ 3.284  \times 10^{  9}$ &0.251&$ 0.000  $                &0.000&0.632&0.131&1.919& 13.39      \\
d&$ 3.319  \times 10^{  9}$ &0.251&$ 0.000  $                &0.000&0.632&0.131&1.919& 13.39      \\
&&&&&&&&&\\
a&$ 1.701  \times 10^{  9}$ &1.496&$ 0.000  $                &0.231&0.700&1.496&1.000& 0.8000     \\
b&$ 3.091  \times 10^{  9}$ &0.825&$-8.864  \times 10^{-10}$ &0.000&0.685&0.494&1.671& 1.022     \\
c&$ 4.056  \times 10^{  9}$ &0.254&$-4.564  \times 10^{-12}$ &0.000&0.641&0.113&2.243& 14.56      \\
d&$ 4.087  \times 10^{  9}$ &0.254&$ 0.000  $                &0.000&0.641&0.113&2.243& 14.56      \\
&&&&&&&&&\\
a&$ 2.162  \times 10^{  9}$ &1.496&$ 0.000  $                &0.000&0.700&1.496&1.000& 1.000      \\
b&$ 2.544  \times 10^{  9}$ &0.890&$-1.750  \times 10^{ -9}$ &0.000&0.685&0.554&1.606& 1.147      \\
c&$ 3.308  \times 10^{  9}$ &0.259&$-1.861  \times 10^{-12}$ &0.000&0.646&0.116&2.237& 17.26      \\
d&$ 3.335  \times 10^{  9}$ &0.259&$ 0.000  $                &0.000&0.646&0.116&2.237& 17.26      \\
&&&&&&&&&\\
a&$ 2.324  \times 10^{  9}$ &1.496&$ 0.000  $                &0.000&0.700&1.496&1.000& 1.300      \\
b&$ 2.437  \times 10^{  9}$ &0.957&$-5.621  \times 10^{ -9}$ &0.000&0.686&0.622&1.539& 1.363      \\
c&$ 2.919  \times 10^{  9}$ &0.265&$-2.269  \times 10^{-12}$ &0.000&0.646&0.119&2.231& 21.12      \\
d&$ 2.942  \times 10^{  9}$ &0.265&$ 0.000  $                &0.000&0.646&0.119&2.231& 21.12      \\
&&&&&&&&&\\
a&$ 1.781  \times 10^{  9}$ &1.496&$ 0.000  $                &0.199&0.700&1.995&0.750& 0.8000    \\
b&$ 3.674  \times 10^{  9}$ &0.723&$-4.551  \times 10^{-10}$ &0.000&0.682&0.474&1.523& 0.8468      \\
c&$ 4.981  \times 10^{  9}$ &0.242&$-1.567  \times 10^{-13}$ &0.000&0.630&0.121&2.004& 9.854      \\
d&$ 2.594  \times 10^{ 10}$ &0.242&$ 0.000  $                &0.000&0.629&0.121&2.004& 9.854      \\
&&&&&&&&&\\
a&$ 2.174  \times 10^{  9}$ &1.496&$ 0.000  $                &0.000&0.700&1.995&0.750& 1.000      \\
b&$ 2.549  \times 10^{  9}$ &0.772&$-8.777  \times 10^{-10}$ &0.000&0.682&0.524&1.474& 0.9587     \\
c&$ 3.593  \times 10^{  9}$ &0.247&$-1.763  \times 10^{-12}$ &0.000&0.635&0.124&1.999& 11.70      \\
d&$ 3.635  \times 10^{  9}$ &0.247&$ 0.000  $                &0.000&0.635&0.124&1.999& 11.70      \\
&&&&&&&&&\\
a&$ 2.347  \times 10^{  9}$ &1.496&$ 0.000  $                &0.000&0.700&1.995&0.750& 1.300      \\
b&$ 2.407  \times 10^{  9}$ &0.824&$-4.119  \times 10^{ -9}$ &0.000&0.682&0.579&1.422& 1.142      \\
c&$ 2.971  \times 10^{  9}$ &0.253&$-7.284  \times 10^{-12}$ &0.000&0.633&0.127&1.993& 14.36      \\
d&$ 3.004  \times 10^{  9}$ &0.253&$ 0.000  $                &0.000&0.633&0.127&1.993& 14.36      \\
&&&&&&&&&\\
a&$ 1.425  \times 10^{  9}$ &1.549&$ 0.000  $                &0.289&0.700&1.496&1.035& 0.8000     \\
b&$ 2.985  \times 10^{  9}$ &0.845&$-1.021  \times 10^{ -9}$ &0.000&0.686&0.486&1.739& 1.039      \\
c&$ 3.841  \times 10^{  9}$ &0.255&$-6.397  \times 10^{-13}$ &0.000&0.638&0.110&2.329& 15.66      \\
d&$ 3.870  \times 10^{  9}$ &0.255&$ 0.000  $                &0.000&0.638&0.110&2.329& 15.66      \\
&&&&&&&&&\\
a&$ 1.890  \times 10^{  9}$ &1.549&$ 0.000  $                &0.057&0.700&1.496&1.035& 1.000      \\
b&$ 2.297  \times 10^{  9}$ &0.903&$-2.169  \times 10^{ -9}$ &0.000&0.686&0.537&1.681& 1.178      \\
c&$ 2.954  \times 10^{  9}$ &0.261&$-1.435  \times 10^{-12}$ &0.000&0.643&0.112&2.323& 18.54      \\
d&$ 2.979  \times 10^{  9}$ &0.261&$ 0.000  $                &0.000&0.643&0.112&2.323& 18.54      \\
&&&&&&&&&\\
a&$ 2.060  \times 10^{  9}$ &1.549&$ 0.000  $                &0.000&0.700&1.496&1.035& 1.300    \\
b&$ 2.180  \times 10^{  9}$ &0.974&$-6.159  \times 10^{ -9}$ &0.000&0.687&0.605&1.610& 1.389        \\
c&$ 2.625  \times 10^{  9}$ &0.267&$-2.597  \times 10^{-12}$ &0.000&0.645&0.115&2.317& 22.68      \\
d&$ 2.646  \times 10^{  9}$ &0.267&$ 0.000  $                &0.000&0.645&0.115&2.317& 22.68      \\
&&&&&&&&&\\
a&$ 1.496  \times 10^{  9}$ &1.549&$ 0.000  $                &0.260&0.700&1.995&0.776& 0.8000    \\
b&$ 3.728  \times 10^{  9}$ &0.734&$-5.001  \times 10^{-10}$ &0.000&0.682&0.461&1.591& 0.8730      \\
c&$ 4.870  \times 10^{  9}$ &0.244&$-8.408  \times 10^{-13}$ &0.000&0.625&0.117&2.081& 10.61      \\
d&$ 4.927  \times 10^{  9}$ &0.244&$ 0.000  $                &0.000&0.625&0.117&2.081& 10.61      \\
&&&&&&&&&\\
a&$ 1.973  \times 10^{  9}$ &1.549&$ 0.000  $                &0.000&0.700&1.995&0.776& 1.000      \\
b&$ 2.251  \times 10^{  9}$ &0.783&$-1.113  \times 10^{ -9}$ &0.000&0.683&0.508&1.542& 0.9872     \\
c&$ 3.117  \times 10^{  9}$ &0.249&$ 0.000  $                &0.000&0.631&0.120&2.076& 12.59      \\
d&$ 3.155  \times 10^{  9}$ &0.249&$ 0.000  $                &0.000&0.631&0.120&2.076& 12.59      \\
&&&&&&&&&\\
a&$ 2.077  \times 10^{  9}$ &1.549&$ 0.000  $                &0.000&0.700&1.995&0.776& 1.300      \\
b&$ 2.150  \times 10^{  9}$ &0.843&$-3.834  \times 10^{ -9}$ &0.000&0.683&0.569&1.482& 1.158     \\
c&$ 2.681  \times 10^{  9}$ &0.255&$ 0.000  $                &0.000&0.631&0.123&2.070& 15.41      \\
d&$ 2.711  \times 10^{  9}$ &0.255&$ 0.000  $                &0.000&0.631&0.123&2.070& 15.41      \\
\hline
\label{8}
\end{tabular}
\end{minipage}
\end{table*}

\end{document}